\documentclass[manuscript]{acmart}
\usepackage{booktabs}
\AtBeginDocument{%
  }

\setcopyright{none}
\usepackage{enumitem}

\usepackage[dvipsnames]{xcolor}
\usepackage{colortbl}
\usepackage{multirow}
\usepackage{array}
\usepackage{siunitx}
\usepackage{graphicx}
\usepackage{ragged2e}
\usepackage{soul}
\usepackage{xspace}
\usepackage{wrapfig}
\usepackage{balance} 
\usepackage{hyperref} 
\usepackage{tabularx} 
\usepackage{adjustbox} 
\usepackage{longtable} 
\usepackage{arydshln}
\definecolor{dashgray}{gray}{0.7}
\usepackage{booktabs}
\usepackage{subcaption}
\usepackage[nameinlink]{cleveref}
\usepackage[normalem]{ulem}
\usepackage{pgfplots}

\usepackage[most]{tcolorbox} 
\tcbset{before skip=10pt, after skip=10pt}

\usepackage{xurl} 
\usepackage{ragged2e} 

\usepackage{dblfloatfix}

\definecolor{navy}{RGB}{0,0,180}
\definecolor{maroon}{RGB}{128,0,0}
\definecolor{forest}{RGB}{34,139,34}

\renewcommand{\emph}[1]{\textit{\underline{#1}}}

\begin{document}

\title[]{Co-Designing AI For Mental Health Support With Young Adults of Color (YOC): Needs, Expectations, and Implications for AI Literacy}


\author{Elaine Dabin Jeon}
\affiliation{%
  \institution{University of Southern California}
  \country{United States}
}

\author{John Bosco S. Bunyi}
\affiliation{%
  \institution{University of California, Riverside}
  \country{United States}
}

\author{Hannah Kim}
\affiliation{%
  \institution{University of Southern California}
  \country{United States}
}

\author{Renkai Ma}
\affiliation{%
  \institution{University of Cincinnati}
  \country{United States}
}

\author{Yaman Yu}
\affiliation{%
  \institution{University of Washington}
  \country{United States}
}

\author{Michal Luria}
\affiliation{%
  \institution{Center for Democracy \& Technology}
  \country{United States}
}

\author{Jason Yip}
\affiliation{%
  \institution{University of Washington}
  \country{United States}
}

\author{Alexis Hiniker}
\affiliation{%
  \institution{University of Washington}
  \country{United States}
}

\author{Katie Davis}
\affiliation{%
  \institution{University of Washington}
  \country{United States}
}

\author{Angel Hsing-Chi Hwang}
\affiliation{%
  \institution{University of Southern California}
  \country{United States}
}

\begin{abstract}
  Young adults of color (YOC) face heightened mental health challenges and barriers to care while navigating developmental and life transitions. Situated between youth-oriented safeguards and adult-oriented AI systems, little is known about how they use AI chatbots for mental health and well-being support or how sociocultural contexts shape their expectations, concerns, and design preferences. We conducted a two-day co-design workshop with 13 Asian, Black, and Hispanic/Latino/a young adults aged 18--24. Participants found generic chatbot advice to flatten their lived experiences; rather than making incorrect assumptions, they wanted more opportunities for identity-informed disclosure. Preferences for YOC-centered personalization also revealed gaps in privacy and AI literacy. Participants negotiated different therapeutic roles for chatbots and sought greater AI accountability and user agency, highlighting blurred boundaries between clinical and non-clinical AI-mediated support. Findings suggest directions for integrating AI literacy with mental health literacy and centering YOC's experiences in the privacy calculus.
\end{abstract}



\authorsaddresses{%
Author's Contact Information: Elaine Dabin Jeon, University of Southern California, United States, elaine.jeon@usc.edu}
\maketitle
\renewcommand{\shortauthors}{Jeon et al.}
\section{Introduction}

Young people are increasingly using consumer-facing AI applications for mental health and well-being support, including general-purpose large language model (LLM)-based chatbots and applications designed for social uses (e.g., AI companions) \cite{mcbain-youth-ai-2025,mcbain-youth-ai-2026,robb-common-sense-media-2025,mcclain-teens-ai-pew-2026}. For \textbf{young adults}---aged 18 to 24 years in this study---the adoption of AI chatbots for such support is unfolding within the broader developmental period of emerging adulthood, characterized by identity exploration, ongoing cognitive development, and major life transitions that can reconfigure social support systems \cite{arnett-2000,arnett-2014,wood-emerging-adulthood-2017}. In the U.S., young adults experience comparatively high rates of mental illness, yet underutilize traditional mental health treatment \cite{nimh-mental-illness}; they also report about twice the rates of anxiety and depression compared to teens \cite{harvard-gse-2023}. Meanwhile, young adults today face many demands and ``less latitude for failure'' \cite{national-academies-2015,harvard-gse-2023}. Against this backdrop, AI chatbots may be perceived as accessible sources of support while also creating potential harms, especially in vulnerable contexts. In this study, we examine the use of AI chatbots for mental health in terms of managing mental illness, as well as emotional, psychological, and social well-being \cite{jung-ai-mh-2025,apa-dictionary}. Such use \cite{torous-digital-mh-2025} ranges from coping with stressors and mental health symptoms (e.g., ADHD, trauma) \cite{luo-ai-mh-2025,jung-ai-mh-2025} to seeking interpersonal advice \cite{tseng-liang-2026}.

For racially and ethnically diverse young adults, these needs can be compounded by heightened stressors \cite{national-academies-2015} and structural barriers to accessing appropriate mental health care \cite{nemoyer-racial-mh-2022,kff-racial-ethnic-disparities-2024}. Recent data, which include ages 18--21, also indicate that they may be using AI chatbots for mental health advice at comparatively higher rates and frequency \cite{mcbain-youth-ai-2026}. Yet, research on conversational AI-mediated mental health support has insufficiently centered cultural and identity-based contexts \cite{wang-ai-mh-2025}. This gap is significant because AI systems are not specifically designed for the critical developmental periods young adults navigate \cite{national-academies-2015}, despite their vulnerability to AI-related harms; those harms are further intensified for young adults of color (YOC) when AI systems reproduce identity-specific biases, stereotypes, and stigma \cite{torous-digital-mh-2025,bouguettaya-racial-bias-2025,gabriel-ai-mh-2024,hofmann-ai-disparities-2024,neplenbroek-2025,pawar-culture-llm-2025}. As cultural backgrounds can also shape attitudes toward AI conversational agents \cite{folk-ai-culture-2025,ge-ai-culture-2024}, YOC's culturally situated knowledge and lived experiences are essential to understanding how they use and make sense of AI chatbots. Accordingly, we collaborated with YOC to examine both \textit{age-related} experiences and \textit{identity- and culturally informed} perceptions and practices surrounding AI use for mental health and well-being support. 

We acknowledge a central challenge: While the risks of using AI chatbots for mental health support are increasingly recognized, discouraging use alone is unlikely to address the needs and circumstances that motivate young adults to seek such support. In addition, AI use for mental health could be further pushed underground, especially since younger users tend not to disclose such use to people in their lives \cite{mcbain-youth-ai-2026}. At the same time, much of AI safety discourse frames ``young people'' primarily as minors, such as age-specific safeguards in commercial platforms like \textit{ChatGPT for Teens} \cite{open-ai-teen}. This framing leaves young adults in an ambiguous position---treated as autonomous adult users despite ongoing psychological and cognitive development and mental health needs. Meanwhile, young adults have unique preferences, attitudes, and utilization patterns compared to the general adult population for chatbot-enabled mental, emotional, and social support \cite{baebrandtzaeg-chatbot-social-support-2021,koulouri-mh-chatbot-2022}. Instead, we build on and extend prior participatory studies that emphasize working \textit{with} people who use AI systems to foster safer, more informed engagement \cite{chang-ai-2025,tian-ai-hallucination-2026,kenny-cail-2025,solyst-youth-ai-auditing-2025,landesman-2026}. This entails identifying young adults' valued design priorities, as well as behavioral and knowledge gaps, to inform AI literacy education that empowers them to navigate their diverse needs. To this end, we ask the following research questions:

\begin{itemize}
    \item \textit{RQ1.} What are YOC's needs, expectations, and concerns regarding AI chatbot use for mental health and well-being support? In what contexts do identity and culture shape these accounts?

    \item \textit{RQ2.} What AI chatbot features do YOC envision for supporting more responsible and mindful mental health-related use?

    \item \textit{RQ3.} What AI literacy implications emerge from YOC's experiences for future mental health contexts?
    
\end{itemize}

We conducted a two-day, in-person co-design workshop with 13 YOC who self-identified as Asian, Black, or Hispanic or Latino/a. We chose co-design and design workshops as an avenue for racially diverse young adults to envision the role of technology in their own futures \cite{harringtonElicitingTechFutures2021}. The workshop combined individual and group reflections, AI conversation exercises, and lightweight, iterative design activities. These activities allowed participants to share contextual knowledge \cite{cao-cail-2025} and personal experiences related to mental health, while collectively reflecting on their understanding of and uncertainties about conversational AI. 

Throughout the workshop, YOC participants expected AI chatbots to provide mental health and well-being support that is more responsive to their sociocultural contexts. They perceived generic chatbot advice as reducing complex concerns---such as parental pressure, mental health stigma, and immigration experience---to typical young-adult ``individual'' problems; however, such advice is often less relevant to challenges connected to YOC's cultural backgrounds and community norms (RQ1). Rather than having systems fill in gaps with incorrect assumptions, participants designed opportunities for identity-informed disclosure (RQ2). For example, they envisioned an initial set of intentional questions (e.g., ethnicity, household dynamics) that would contextualize subsequent conversations. However, with the desire to disclose more of their personal backgrounds, participants rarely discussed the privacy implications of in-depth self-disclosure (RQ3).

Additionally, participants wanted chatbots to clearly communicate the forms of mental health support they can or cannot provide from the outset. They envisioned chatbots as serving multiple functions across the help-seeking process, including culturally resonant therapeutic roles (RQ1). Their designs foregrounded features intended to strengthen AI accountability and user agency (RQ2). At the same time, participants struggled to reach consensus about the appropriate role of AI conversational agents within the formal mental health support landscape and to grasp what system ``pushbacks'' and safeguards look like (RQ3). Drawing on these findings, we conceptualize how gaps in AI literacy may amplify existing gaps in mental health literacy, motivating a more integrated approach to AI literacy for young adults engaging with AI-mediated mental health and well-being support.

This paper makes three contributions: 
\begin{enumerate}
    \item \textit{An empirical account of how YOC's identities, lived experiences, and mental models of AI shape their uses and perceptions of AI chatbots for mental health and well-being support.} Drawing on participatory design approaches, we contribute to a more population- and domain-specific account of AI use, safety, and literacy that moves beyond individualistic models of AI adoption and use. 
     \item \textit{A set of design principles for AI chatbot features that support YOC's vision of mindful and culturally responsive engagement with AI chatbots for mental health and well-being.} These principles help identify opportunities to extend broad AI safety approaches by attending to more identity-specific vulnerabilities. 
      \item \textit{A conceptual framework for AI literacy, mental health literacy, and privacy considerations.} We show why informed and safer engagement with AI chatbots for mental health support cannot be reduced to technical knowledge or prompting skills alone; it also requires capacities to navigate AI chatbots' structural affordances and relational dynamics, critically assess their limitations and harms in mental health contexts, and recognize culturally situated risks. 
\end{enumerate}

\section{Background and Related Work}

\subsection{Young Adults' AI Chatbot Use for Mental Health Support and Its Limitations}
As more and more young adults, especially YOC, turn to AI chatbots for mental health and well-being support, their wide-ranging perspectives and practices should be further explored. This is pressing given that AI chatbots are rapidly evolving and being adopted at a pace that currently exceeds our understanding of their appropriateness and effectiveness for mental health support. Further, much more should be done to translate these experiences into designing technologies and evaluating the resulting impact.

\subsubsection{Use of Conversational AI for Mental Health and Well-Being Support}

``Mental health'' encompasses both mental illness and subjective emotional, psychological, and social well-being, including people's capacity to manage stressors and navigate relationships \cite{jung-ai-mh-2025,apa-dictionary}. Consistent with this broad conceptualization, a recent review including participants aged 18--24 found that they rarely hold a single, shared definition of mental health; rather, they understood mental health to cover emotions, coping, relationships, identity, and life satisfaction \cite{beckman-children-2023}. Given this, mental health-related conversations with AI chatbots span concerns with different meanings and risk profiles---affective or interpersonal concerns (e.g., navigating relationships \cite{tseng-liang-2026}), clinically framed help-seeking (e.g., managing mental health conditions \cite{luo-ai-mh-2025}), and high-risk or crisis conversations---and accordingly require different forms of monitoring and safeguards \cite{mcbain-ai-mh-convo-2026}. 

Young adults may turn to AI chatbots for diverse mental health concerns, particularly given they already view AI-enabled systems as a potential means of navigating gaps in the help-seeking process \cite{petersson-ai-ya-mh-2025}. However, less is known about \textit{how} young adults specifically use and perceive general-purpose and consumer-facing AI chatbots for such concerns \cite{daswani-chatgpt-mh-2026}. Young adulthood entails distinctive developmental and social circumstances \cite{arnett-2000,national-academies-2015}, including patterns of support-seeking outside traditional sources that often intersect with digital technology use \cite{meyerhoff-young-adults-2022,kauer-young-people-dmh-2014}. Together, these contexts may shape young adults' needs and expectations for mental, emotional, and social support from conversational AI in ways that differ from those of older adults and younger teens \cite{baebrandtzaeg-chatbot-social-support-2021,koulouri-mh-chatbot-2022}. For instance, they report greater familiarity with and use of AI than other-age adults \cite{kennedy-pew-age-ai-2025}. Further, among surveyed teens and young adults, respondents aged 18--21 had greater and increasing use of AI chatbots for mental health advice relative to younger respondents \cite{mcbain-youth-ai-2025,mcbain-youth-ai-2026}.


This gap is especially salient for YOC. Available evidence suggests that racially and ethnically diverse teens and young adults report high and frequent use of AI chatbots for mental health support \cite{mcclain-ai-pew-2026,mcbain-youth-ai-2026}, yet YOC's specific practices, perceptions, and strategies for navigating chatbot-mediated support remain poorly understood. Furthermore, cultural contexts can shape users' attitudes toward conversational AI and their expectations of these systems, including preferences for AI autonomy and relational connection \cite{folk-ai-culture-2025,ge-ai-culture-2024}. Further research is therefore needed to examine how YOC's diverse, culturally situated experiences shape mental health-related interactions with AI chatbots.  

\subsubsection{Emerging Research on Limitations of AI-Mediated Support}
The growing use of general-purpose LLM chatbots for mental health support is concerning because these systems are not designed or clinically validated to provide therapeutic care \cite{yee-stanford-2026}. Documented limitations and harms include inadequate detection and protections in high-risk or crisis situations, with increasing attention on younger users \cite{duffy-cnn-2024,chatterjee-npr-2025,kuznia-cnn-2025}. Further, when users bring clinically framed concerns to AI chatbots, these systems can respond with stigma \cite{moore-stigma-llm-2025}, provide unauthorized medical advice \cite{li-counselbench-2026}, and respond inappropriately to common symptoms \cite{moore-stigma-llm-2025}. Even in affective or interpersonal conversations, risks can arise through structural components, such as data collection and privacy practices, and relational dynamics, such as emotional reliance \cite{apa-ai-advisory,hiniker-deceptive-2026,mcbain-ai-mh-convo-2026}. 

However, limited research examines young adults as a distinct developmental group in defining and evaluating AI safety. Emerging research focused on those under 18 illustrates the importance of developmentally informed safety approaches. For example, the Youth-Centered GAI Risks (YAIR) taxonomy identifies risks—including risks to mental well-being and social development—that existing AI safety frameworks do not fully capture \cite{yu-yair-2025}. Yet, it is unclear the extent to which youth-centered safety frameworks extend to young adulthood in light of the distinctive forms of autonomy, identity exploration, changing social networks, digital support-seeking, and life transitions---on top of ongoing cognitive development \cite{arnett-2000,arnett-2014,wood-emerging-adulthood-2017,national-academies-2015,meyerhoff-young-adults-2022}. More work should focus on building upon developmentally informed approaches, especially in vulnerable contexts like mental health support-seeking. Such work must also account for the experiences of YOC, for whom racial bias, stereotypes, and stigma reproduced or amplified by AI systems \cite{torous-digital-mh-2025,bouguettaya-racial-bias-2025,gabriel-ai-mh-2024,hofmann-ai-disparities-2024,neplenbroek-2025,pawar-culture-llm-2025} may compound existing inequities in mental health contexts \cite{kff-racial-ethnic-disparities-2024,alegria-mh-disparities-2010}.

Given these concerns, studies have begun to design interventions intended to promote more mindful and safer AI chatbot use, though evidence of their effectiveness remains limited. Examples include warning labels and other safety-oriented interface features \cite{tian-ai-hallucination-2026,ibrahim-ai-sycophancy-2026,krsek-disclosure-2026,poonsiriwong-death-ai-2026}. In addition to the lack of evidence of their effectiveness, existing interventions and platform safeguards have rarely been designed with, or rigorously evaluated among, young adults---particularly YOC. This concern echoes the principle ``nothing about us without us'' in disability justice and design scholarship, which holds that those most affected by a design decision must be included in \textit{shaping} it rather than simply consulted after the fact \cite{charlton-disability-1998,costanza-chock-design-justice-2020}. Without YOC's participation, efforts to make AI chatbots ``safer'' risk reproducing existing assumptions and harms that have already unequally impacted their experiences in mental health and technology design. 

\subsection{Approaches to Supporting AI Use for Mental Health among Young Adults}
Beyond addressing gaps in YOC-centered design of AI chatbots, we argue that it is also important to understand whether young adults have the AI literacy needed to use conversational AI mindfully and responsibly for mental health and well-being support. Such literacy empowers users to recognize how AI systems work, and this work starts with reviewing their current limitations and making AI literacy contextually appropriate. Prior work has generally pursued two approaches: (1) developing AI literacy frameworks and programs, and (2) engaging young adults through participatory design methods.

\subsubsection{Existing AI Literacy Frameworks for Young Adults} 

Over the past decade, AI literacy scholarship has expanded alongside AI's growing integration into everyday life, identifying the knowledge and competencies needed to help people understand, evaluate, and use AI systems \cite{long-magerko-2020,ng-ai-literacy-2021}. Early work articulated foundational concepts, including five ``big ideas'' for AI education \cite{touretzky-ai-2019}, and defined AI literacy as competencies needed for critically evaluating AI, communicating and collaborating with AI systems, and using them across varied contexts \cite{long-magerko-2020}. With the widespread availability of generative AI, more recent frameworks have incorporated practical competencies (e.g., prompting and detecting AI-generated content) alongside sociotechnical and ethical awareness \cite{annapureddy-ai-2025,jia-ai-2025}. 

Nevertheless, it remains unclear how well existing AI literacy education equips young adults to navigate conversational AI in personal and higher-stakes contexts, including mental health and well-being support. Research on AI literacy has mainly focused on K--12 \cite{touretzky-ai-2019,wong-ai-literacy-2020,casal-ai-2023} or higher-education settings \cite{gu-ai-2025,laupichler-ai-literacy-2022}. Although these efforts provide an important foundation, young adults specifically may encounter AI literacy unevenly---some in K--12 or university settings, whereas others may have limited access to formal AI literacy education or no longer be situated in educational institutions. Moreover, even when available, classroom-based approaches may not address the personal, relational, and emotionally consequential uses of conversational AI that occur outside the academic or professional uses \cite{xie-ai-roles-2025}. 

A further limitation of conventional AI literacy education concerns the distinctive nature of mental health support-seeking with conversational agents. Existing AI literacy competencies may help users understand AI capabilities and limitations or identify bias and misinformation \cite{touretzky-ai-2019,annapureddy-ai-2025}; however, they may be insufficient for recognizing interactional risks that emerge over time. For example, recognizing harmful language in chatbot output does not necessarily enable users to identify emotional dependence, escalating reliance, or blurred boundaries across repeated interactions with an AI chatbot \cite{yu-yair-2025}. Emerging K--12-focused initiatives begin to connect AI literacy with well-being \cite{csm-curriculum}, though their applicability to young adults and to mental health-related chatbot use remains unclear.

These gaps may be further heightened for YOC, who encounter racialized and culturally misaligned biases in AI systems \cite{hofmann-ai-disparities-2024,gabriel-ai-mh-2024,neplenbroek-2025,pawar-culture-llm-2025}, yet report comparatively high use of AI chatbots for mental health advice \cite{mcbain-youth-ai-2026}. Their interactions with chatbots may also be shaped by diverse sociocultural contexts and culturally situated understandings of mental health that remain under-examined. More research is needed to identify the knowledge and behavioral gaps that YOC experience while using conversational AI for mental health and well-being support, including how these needs vary across sociocultural contexts and patterns of use.

\subsubsection{Participatory Design Methods for AI Literacy and Young Adults}

AI literacy research increasingly draws on participatory design to foreground people's situated knowledge and everyday experiences of AI, in topics such as AI fairness and critical AI literacy. Emerging studies have, for example, extended this approach to teens \cite{solyst-youth-ai-auditing-2025,chang-ai-2025,kenny-cail-2025,landesman-2026}. Parallel work in digital mental health has applied co-design methods with young adults to identify their needs and preferences for technology-mediated mental health support \cite{figueroa-2026,chinsen-review-2025,mcgovern-review-2025}. However, limited research brings these bodies of work together by engaging \textit{specifically young adults} as active contributors to the design of AI literacy. This gap stands out given limited attention to demographic factors---for example, ethnicity and socioeconomic backgrounds---within digital mental health research that includes young adults \cite{bergin-dhmi-review2020}. It also highlights the need to move beyond designing for young adults toward designing with them with more empathy-driven design \cite{scholten-design-2019}. Building on these works, this study takes a participatory design approach, and the Methods section further details how and why this approach was operationalized.

\section{Methods}
We conducted a two-day, in-person co-design workshop with young adults who self-identified as Asian, Black, or Hispanic or Latino/a (\textit{n} = 13, aged 18–24, \textit{M} = 20, \textit{SD} = 2.29) and five facilitators. Co-design entails a process of ``joint inquiry and imagination'' \cite{steen-co-design-2013}, grounded in participatory design commitments to mutual learning and shared power and decision-making \cite{robertson-mutual-learning-2014}. For young adults,  participatory design positions participants as co-constituents---rather than merely subjects---of mental health research that intersects with digital technology, given they are digital natives and value agency \cite{scholten-design-2019,mcgovern-review-2025}. Accordingly, we used co-design to extend participatory approaches in the context of AI-mediated mental health and well-being support. Our approach also builds on prior work that has centered young people, such as teens, as active problem-solvers in AI knowledge production \cite{solyst-youth-ai-auditing-2025,chang-ai-2025,kenny-cail-2025,landesman-2026}.

We developed the workshop as a pilot model for engaging YOC in conversations about AI chatbots and mental health. Given the topic's sensitivity, we first conducted the workshop with young adults and plan to adapt the approach for other populations, such as younger teens and individuals living with mental health diagnoses. We operationalized co-design through workshops, a common HCI format for eliciting participants' values, needs, and non-design expertise around emerging technologies \cite{haghighi-2023}. Our workshop incorporated lightweight and iterative design activities, such as storyboarding and paper-based prototyping. We also incorporated activities and facilitation practices intended to foster psychological safety, build trust, and support individual reflection to help minimize emotional burden on participants.

We pilot-tested the workshop with three young adults outside the participant pool. We also iteratively reviewed and refined the workshop program and materials with a panel of six experts with extensive experience conducting youth-centered design research. The study was approved by the University's Institutional Review Board. 

\subsection{Participants}
We recruited participants through previous Youth Participatory Action Research \cite{khawaja-ypar-2024} programs, community organizations, and word of mouth. Participants were eligible if they: (1) were 18-24 years old; (2) self-identified as a person of color (defined as identifying with a non-White race); and (3) used AI chatbots regularly for any purpose (defined as at least once a week). We did not restrict candidates from joining if they had never used an AI chatbot for mental health and well-being support to accommodate a broad range of perspectives and attitudes. The workshop took place during Summer 2026 in a major metropolitan area on the West Coast of the U.S. with a racially and culturally diverse population. Of 29 people who expressed interest, 13 were invited to participate based on eligibility, availability to join in-person, and complete interest form.

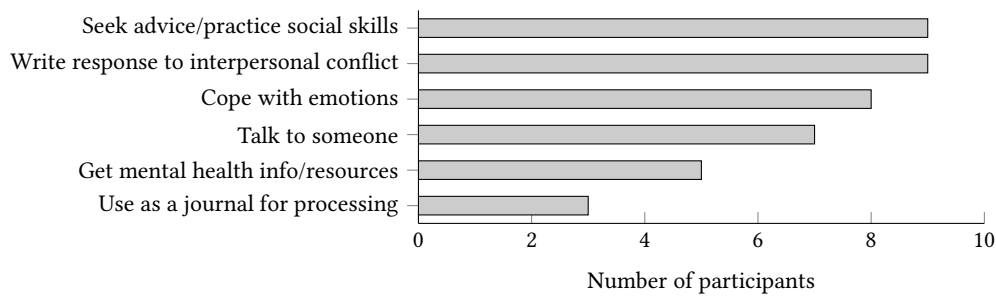
\begin{figure}
    \centering
    \begin{tikzpicture}
        \begin{axis}[
            xbar,
            xmin=0,
            xmax=10,
            axis y line*=left,
            axis x line*=bottom,
            width=0.6\textwidth,
            height=4.4cm, 
            xlabel={Number of participants},
            symbolic y coords={
                Seek advice/practice social skills,
                Write response to interpersonal conflict,
                Cope with emotions,
                Talk to someone,
                Get mental health info/resources,
                Use as a journal for processing
            },
            ytick=data,
            y dir=reverse,
            bar width=7pt, 
            cycle list={}
        ]
            \addplot[draw=black, fill=gray!40] coordinates {
                (9,Seek advice/practice social skills)
                (9,Write response to interpersonal conflict)
                (8,Cope with emotions)
                (7,Talk to someone)
                (5,Get mental health info/resources)
                (3,Use as a journal for processing)
            };
        \end{axis}
    \end{tikzpicture}
    \caption{AI chatbot use cases for mental health and well-being support among participants (\textit{n =} 13) (multiple response options allowed)}
    \label{fig:ai-mh_uses}
\end{figure}

Most participants reported using AI chatbots for \textit{any purposes} either daily/almost-daily (53.8\%, \textit{n =} 7) or at least once a week (46.2\%, \textit{n =} 6). ChatGPT and Gemini were the most commonly used LLM-based chatbots. The use of AI chatbots for \textit{mental health and well-being support} was less frequent: one (7.7\%), three (23.1\%), and five (38.4\%) participants reported mental health-related AI chatbot use daily/almost-daily, at least once a week, and at least once a month, respectively~(\Cref{fig:ai-mh_uses}). Twelve participants attended both workshop days; one participant who could only attend Day 1 completed a one-hour follow-up interview. Participants who attended both days received \$180. Before Day 1 activities started, the first author explained the study procedures, including data collection, to the participants in person. Participants then provided their written informed consent by signing physical forms. 

\begin{table}
  \centering
  \small
  \caption{Participant characteristics (\textit{n} = 13)}
  \label{tab:participant-demographics}
  \begin{tabular}{@{}lllccl@{}}
    \toprule
    ID & Race & Gender & Age & \begin{tabular}[b]{@{}c@{}}Perceived\\SES\end{tabular} & \begin{tabular}[b]{@{}l@{}}AI chatbot use for\\mental health\end{tabular} \\
    \midrule
    P01 & African American / Black & Man & 20 & 3 & Never \\
    P02 & African American / Black & Woman & 20 & 1 & At least once a week \\
    P03 & Asian / Asian American & Man & 20 & 3 & At least once a month \\
    P04 & Asian / Asian American & Woman & 20 & 3 & At least once a month \\
    P05 & Asian / Asian American & Woman & 23 & 3 & At least once a week \\
    P06 & Asian / Asian American & Woman & 24 & 5 & Daily or almost daily \\
    P07 & Asian / Asian American & Woman & 24 & 3 & At least once a week \\
    P08 & Asian / Asian American & Woman & 18 & 3 & Hardly ever \\
    P09 & Asian / Asian American & Woman & 18 & 2 & At least once a month \\
    P10 & Hispanic / Latino/a & Man & 24 & 2 & Hardly ever \\
    P11 & Hispanic / Latino/a & Woman & 18 & 1 & At least once a month \\
    P12 & Hispanic / Latino/a & Woman & 20 & 2 & At least once a month \\
    P13 & Hispanic / Latino/a & Woman & 20 & 2 & Hardly ever \\
    \bottomrule
  \end{tabular}

  \vspace{2pt}
  {\footnotesize\raggedright
  \textit{Note.} Perceived SES denotes participants' perception of their socioeconomic status. 1 = not enough to get by, 2 = just enough to get by, 3 = we only have to worry about money for fun or extras, 4 = we never have to worry about money, 5 = it varies a lot (e.g., based on seasonal sources of income).\par}
\end{table}

\begin{figure}
    \centering
    \includegraphics[width=1\linewidth]{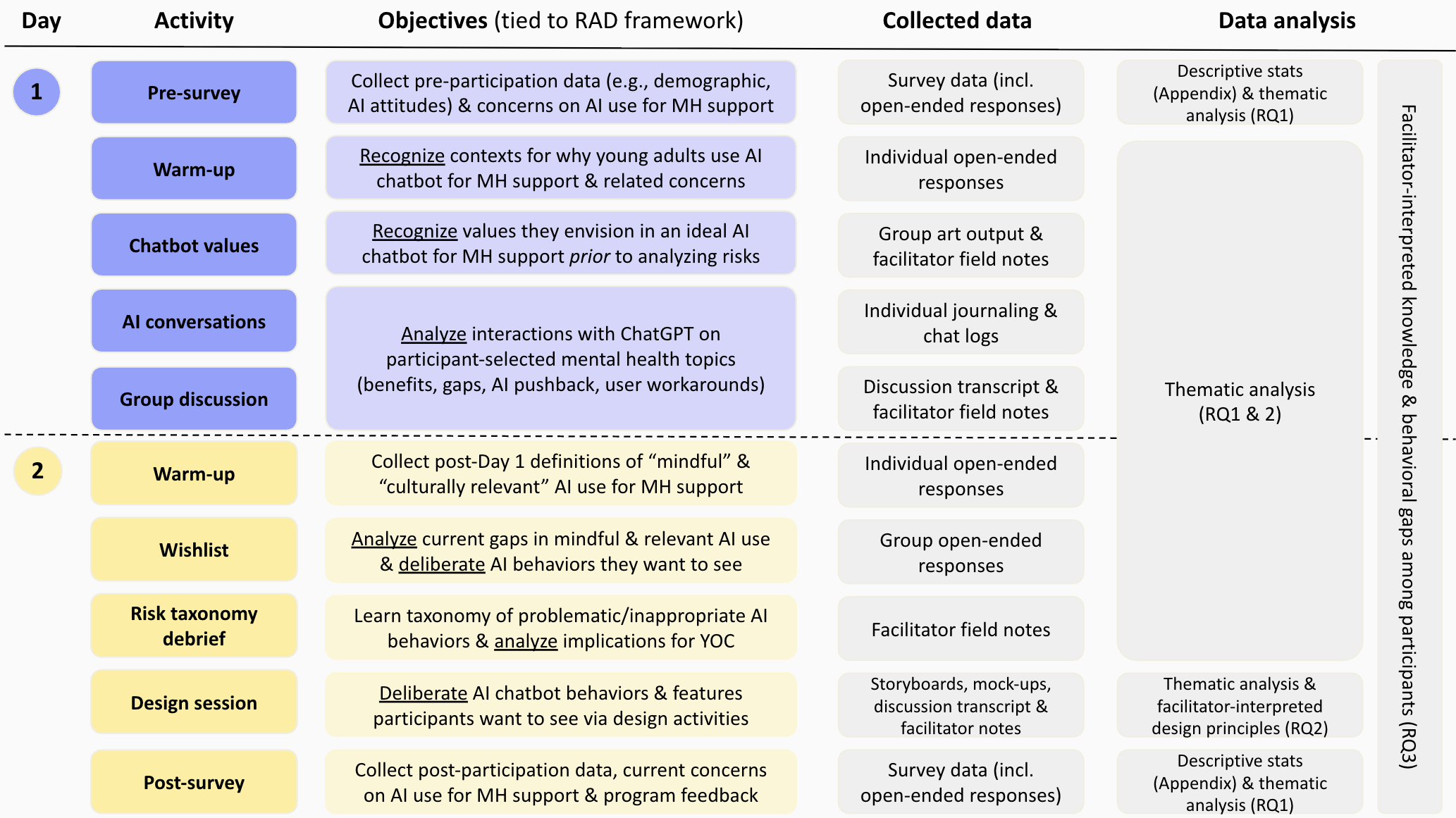}
    \caption{Overview of two-day workshop activities, objectives, collected data, and data analysis methods}
    \label{fig:workshop-diagram}
\end{figure}

\subsection{Day 1 – Recognize and Analyze: Reflecting on AI Chatbot Use for Mental Health}
Our workshop design draws on Solyst et al. (2025)'s RAD framework, which guides participants through three stages of critically examining AI: \textit{Recognize} (identifying potential AI harms), \textit{Analyze} (understanding sociotechnical dimensions of harms), and \textit{Deliberate} (developing recommendations for improvement) \cite{solyst-rad-2025}~(\Cref{fig:workshop-diagram}). We adapted this approach for our YOC participants, given their varying levels of experience and familiarity with conversational AI for mental health and well-being support. 

Day 1's activities created opportunities for participants to \textit{recognize} their own uses of AI chatbots, share perspectives with peers, and identify initial concerns, needs, and expectations related to chatbot use for mental health and well-being support (RQ1). These activities prepared participants to transition to \textit{analyzing} ChatGPT interactions for mental health support more critically. We postponed any informational session on problematic AI behaviors until Day 2. This sequencing was intended to avoid prematurely framing participants' interpretations.

\subsubsection{Warm-up and Group Reflection on Ideal AI Chatbot for Mental Health}
Day 1 began with participants completing a pre-participation survey and generating their definitions of mental health and well-being. Rather than imposing a predefined clinical or expert definition, we encouraged a broad conceptualization \cite{beckman-children-2023} from participants. Responses included \textit{mood regulation, stable emotions, low distress, inner peace, self-confidence, life quality,} and \textit{holistic health.} Participants then discussed why they or their peers might seek mental health and well-being support from AI chatbots, as well as their initial concerns about this form of AI use.

After the warm-up, participants worked in groups on an art-based activity informed by value-sensitive design \cite{friedman-value-sensitive-1996}. They were asked to consider the following prompt: “If a friend were going to use an AI chatbot for mental health, what would you hope the chatbot does well?” Groups identified values they hoped such a chatbot would embody, then brainstormed an object, space (e.g., a room), or character that could represent those values. Each group drew one example and the values it reflected. This activity served as a throughline for the remainder of the workshop: participants frequently returned to the values they identified during this early exercise when navigating AI conversations and design sessions. Final examples included a garden, a library with a lighthouse, and a scale.

\begin{figure}[htbp]
    \centering
    \begin{subfigure}[b]{0.48\textwidth}
        \centering
        \includegraphics[width=\textwidth]{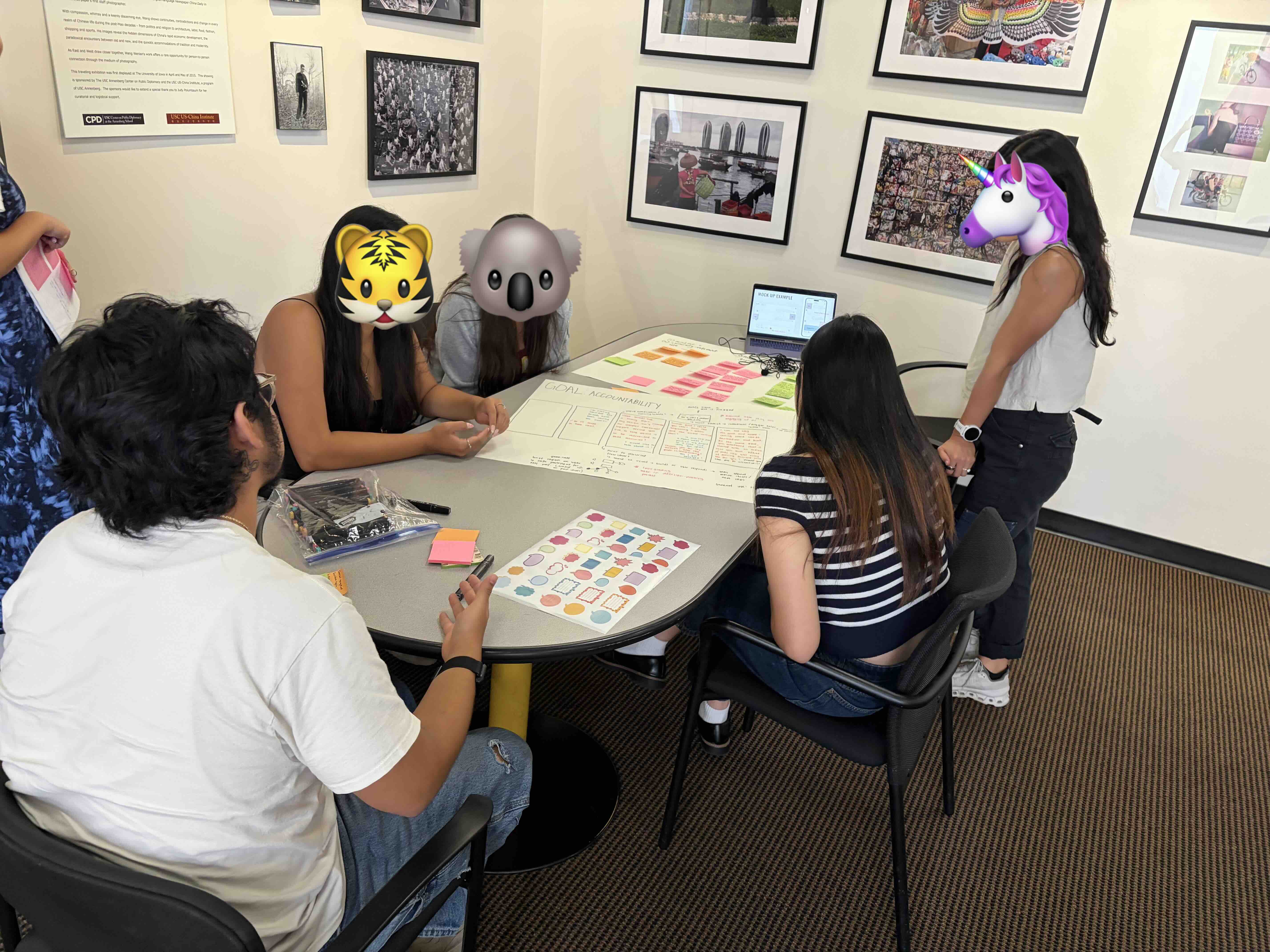}
    \end{subfigure}
    \hfill
    \begin{subfigure}[b]{0.48\textwidth}
        \centering
        \includegraphics[width=\textwidth]{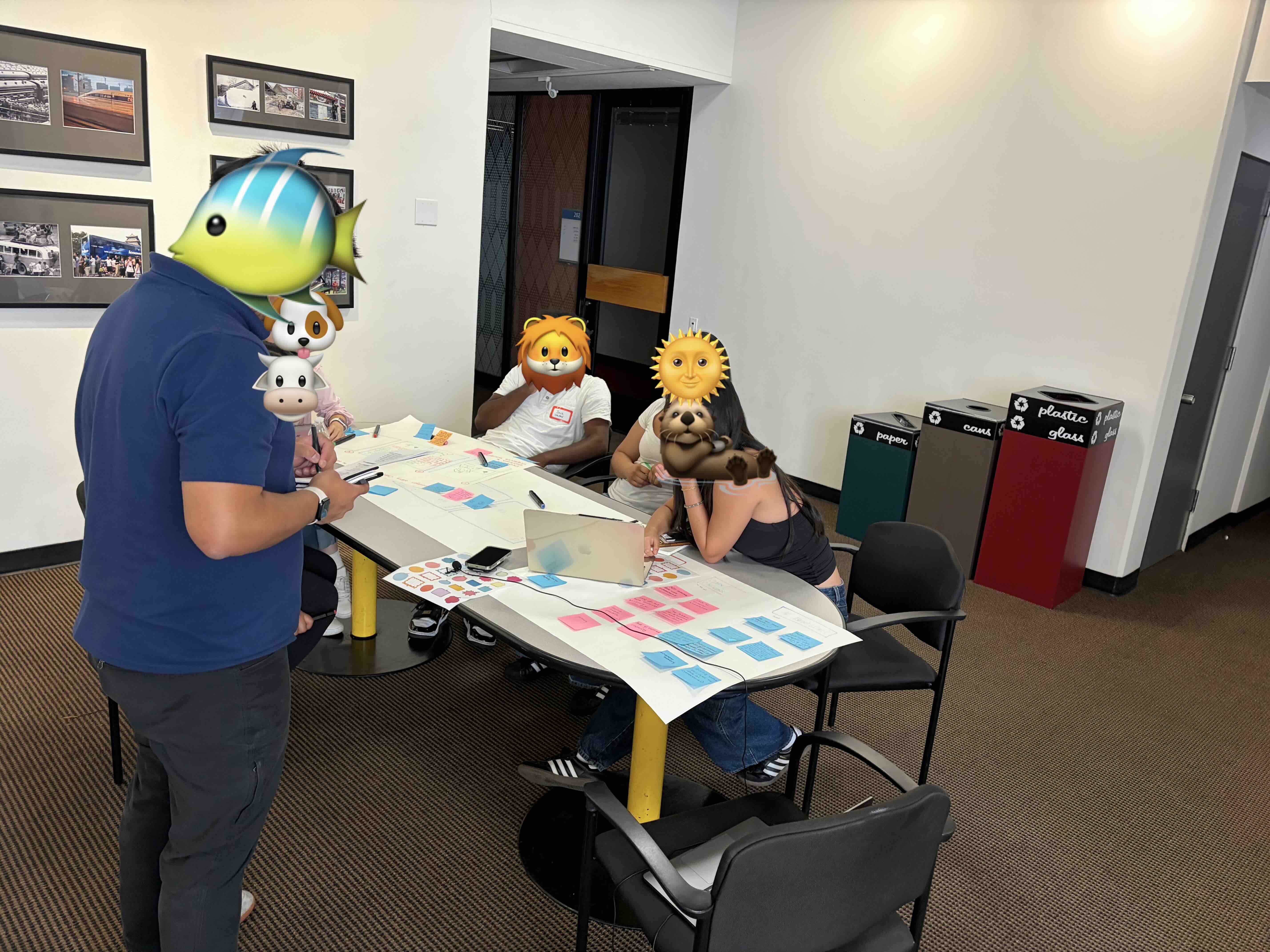}
    \end{subfigure}
    \caption{Throughout the two-day workshop, participants engaged in several small group activities to discuss, brainstorm, and co-design their desirable forms of conversational AI for mental health and well-being support.}
    \label{fig:groupwork}
\end{figure}

\subsubsection{AI Chatbot Conversations, Journaling, and Group Discussion}
Participants engaged in two 15-minute conversations with ChatGPT\footnote{To protect participants' privacy and data, all participants used research team members' University-contracted ChatGPT EDU accounts, which do not store nor use user data for model training. During the chat sessions, they also turned on the temporary chat mode for privacy and data protection.}, individually or in pairs depending on their preference. In the first conversation, participants responded to: “What would another young person of color ask an AI chatbot for their mental health and well-being? Feel free to draw on your identity and backgrounds to inform your prompts.” They explored topics including social anxiety, stress and workload management, parental pressure, interpersonal conflict, and immigration experience. In the second conversation, they considered when an AI chatbot might push back in a mental health context and how users might still attempt to elicit cooperation; they explored scenarios such as asking ChatGPT to act as a therapist (e.g., diagnose symptoms, conduct CBT), seeking validation for cheating on an exam, and navigating moral ambiguity in relationships. 

Before the activity, we reviewed safety guidelines, including avoiding illegal or harmful topics and emphasizing that ChatGPT is not a clinically trained provider. After each conversation,  participants wrote a brief product-style review of their interaction as part of individual journaling (Appendix~\ref{app:journaling}) and consented to donate their in-session chat log data. Day 1 concluded with a semi-structured group discussion of participants' observations and reflections (Appendix~\ref{app:groupq}). 

\subsection{Day 2  – Analyze and Deliberate: Translating Gaps into Design Solutions}

Day 2 dedicated more time to \textit{analyzing} current gaps and potential harms that hinder mindful and relevant use of AI chatbots for mental health and well-being support and transitioned to \textit{deliberating} what AI chatbot behaviors and features participants wanted to see (RQ2). We underscored ideation, rather than perfection, in group debriefs and design sessions, so that we could use participants' discussions to surface knowledge and behavioral gaps to highlight AI literacy implications (RQ3).

\subsubsection{Wishlist Brainstorming and Taxonomy of Problematic AI Behaviors}
\begin{figure}[htbp]
    \centering
    \begin{subfigure}[b]{0.48\textwidth}
        \centering
        \includegraphics[width=\textwidth]{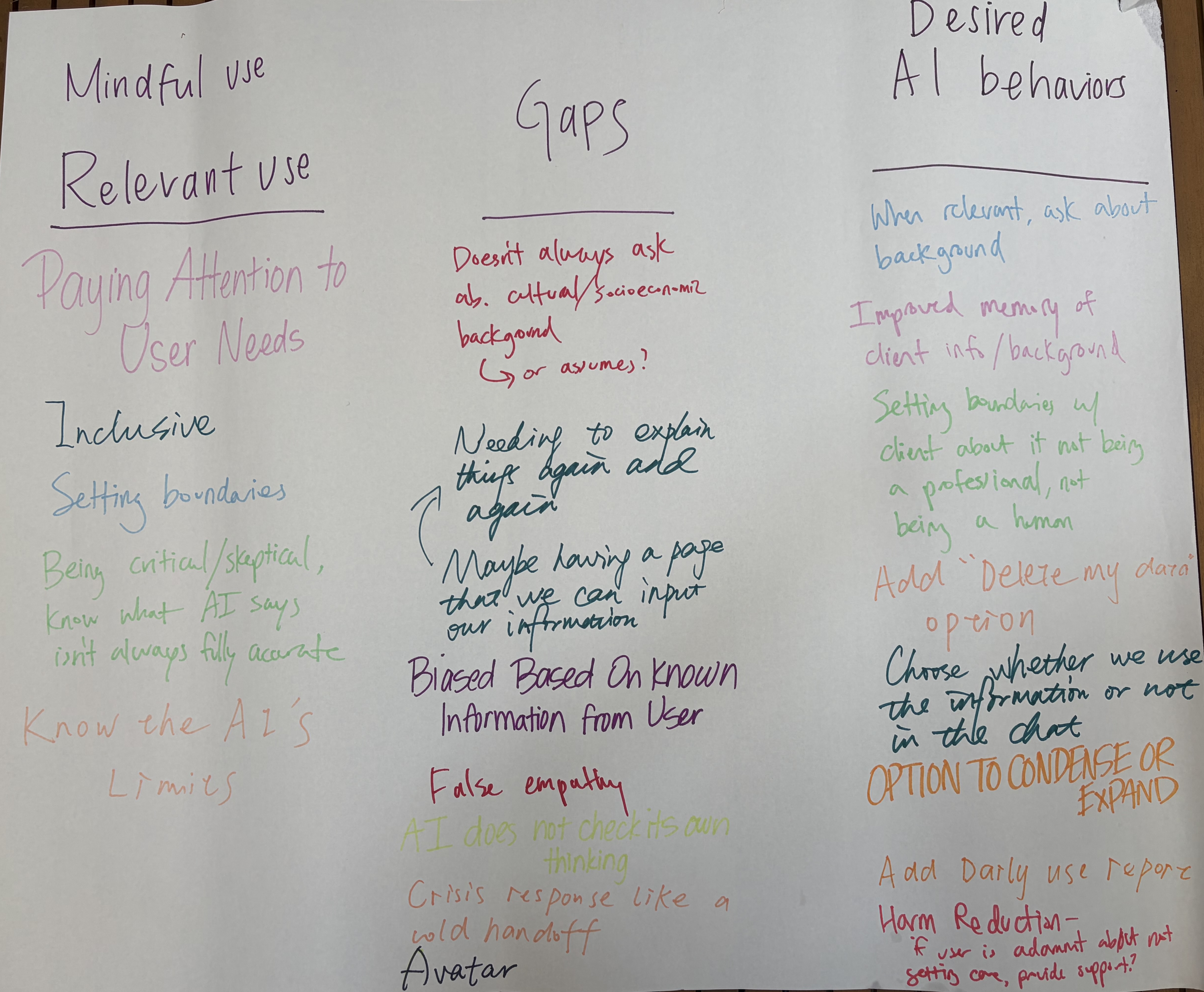}
    \end{subfigure}
    \hfill
    \begin{subfigure}[b]{0.48\textwidth}
        \centering
        \includegraphics[width=\textwidth]{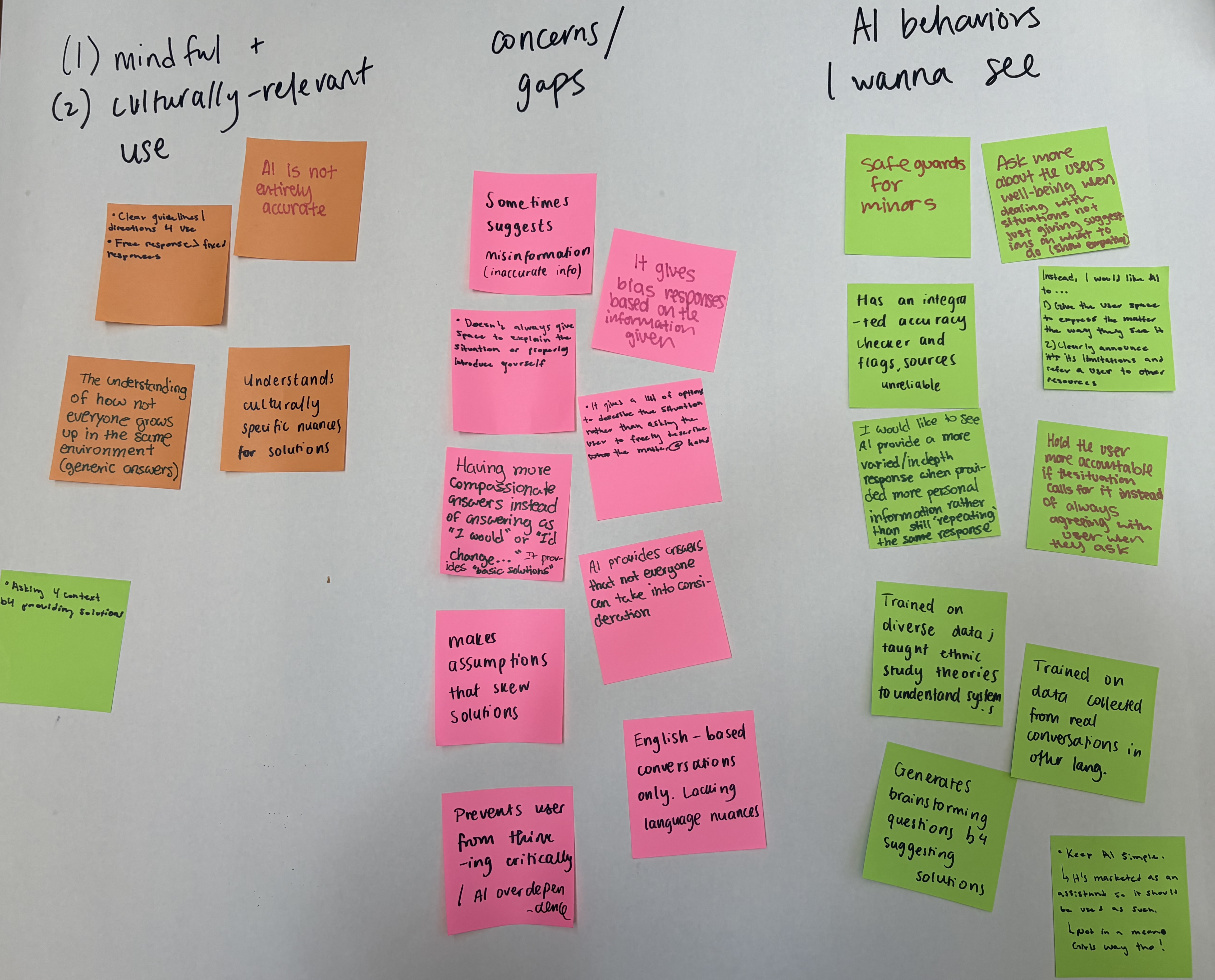}
    \end{subfigure}
    \caption{Day 2 activity identifying values in AI chatbot, gaps or concerns, and wishlist of AI behaviors}
    \label{fig:day2_wishlist}
\end{figure}

Building on Day 1 reflections, participants started Day 2 by collaboratively brainstorming a “wishlist” of AI chatbot behaviors to support mindful and culturally relevant mental health and well-being use. They defined these concepts in their own terms, identified current pain points, and proposed desired chatbot behaviors (Figure~\ref{fig:day2_wishlist}). After initial brainstorming, we introduced a taxonomy of potentially problematic or inappropriate AI behaviors for young adults in mental health contexts (Appendix~\ref{app:taxonomy}), developed from American Psychological Association's guidance on AI-generated mental health advice \cite{apa-ai-advisory} and a targeted literature review. We refined the taxonomy through consultation with AI safety researchers, clinical experts, and non-participating young adults, prioritizing observable interactional behaviors and structural factors salient to mental and emotional support and to YOC. Participants debriefed after the taxonomy overview and finalized their wishlists.

\subsubsection{Co-design of AI Behaviors and Solutions}
We introduced storyboarding and low-fidelity prototyping to scaffold participants' design process. Four teams selected an AI behavior from their wishlists and developed a storyboard around a self-selected scenario (e.g., “young adults asking ChatGPT for advice on how to come out to their family”). Teams then selected a user goal (e.g., “process emotions and organize thoughts”) for the storyboard's first panel and translated their storyboards into low-fidelity mock-ups. An undergraduate student with extensive UI/UX experience served as a design consultant, supporting participants' visualization of their ideas and, when appropriate, using generative AI tools to guide this process. The workshop concluded with a presentation of and peer feedback on the mock-ups (Appendix~\ref{app:design-outputs}), and a post-participation survey.

\subsection{Data Collection and Analysis}
We collected multiple forms of workshop data. Four facilitators audio-recorded group discussions and took extensive field notes during and immediately after workshop activities. We also collected individual responses from warm-up activities, journaling exercises, donated in-session chat logs, and survey responses.

We conducted an inductive thematic analysis of these data sources, following Braun and Clarke's six-phase process \cite{braun-clarke-2006}. An inductive approach enabled us to better foreground participants' own perspectives on using conversational AI for mental health and their desired AI behaviors. Furthermore, there was no established theoretical or design framework to adequately and consistently guide a deductive analysis of YOC's experiences with AI chatbots for mental health. At the same time, our research questions, workshop activities, and facilitators' follow-up prompts shaped the topics participants were invited to discuss and, consequently, the data available for analysis. We therefore treated participants' accounts as situated within the workshop context rather than as exhaustive representations of their lived experiences.

Throughout and following the workshop, the facilitators engaged in iterative discussions of preliminary observations and field notes. Two authors then familiarized themselves with the data by reviewing field notes, audio recordings, and all other data sources and generated initial codes inductively, which were then iteratively organized and refined into themes and subthemes. They met weekly to discuss emerging findings from the data. To mitigate potential biases, they continuously shared and critically examined their interpretations with the broader research team. Still, we acknowledge that authors' lived experiences and professional and educational backgrounds can shape interpretation of the data. We provide a positionality statement in Appendix~\ref{app:positionality} to contextualize how the identities and backgrounds of the research team shaped data collection and interpretation. 

We also collected participant responses on AI attitudes and literacy from pre- and post-participation surveys (\textit{n = }12 with two-day participation). We detail the survey measures and report these data in Appendix~\ref{app:survey}.

\section{Findings}
We organize our findings around two themes. First, participants found generic mental health advice and incorrect assumptions from AI chatbots reduce their lived experiences and concerns to problems expected of typical young adults; instead, they designed for more \textit{identity-informed disclosure}~(\Cref{findings-QA}). Second, participants negotiated the \textit{roles} they expected chatbots to perform in the help-seeking process, including culturally resonant therapeutic roles, and designed for greater AI accountability and user agency in the chatbot designs ~(\Cref{findings-role}). We report findings shaped by YOC's identities and sociocultural contexts, alongside findings that reflect more age-specific experiences elicited through participant-generated prompts (e.g., academic performance, internship). For each theme, we organize our findings around participants’ current pain points and desired alternatives for AI use for mental health and well-being support (RQ1), their designs for addressing these considerations (RQ2), and remaining knowledge or behavioral gaps that point to limitations in AI literacy in mental health contexts (RQ3).

\begin{figure}[h!]
    \centering
    \includegraphics[width=0.85\linewidth]{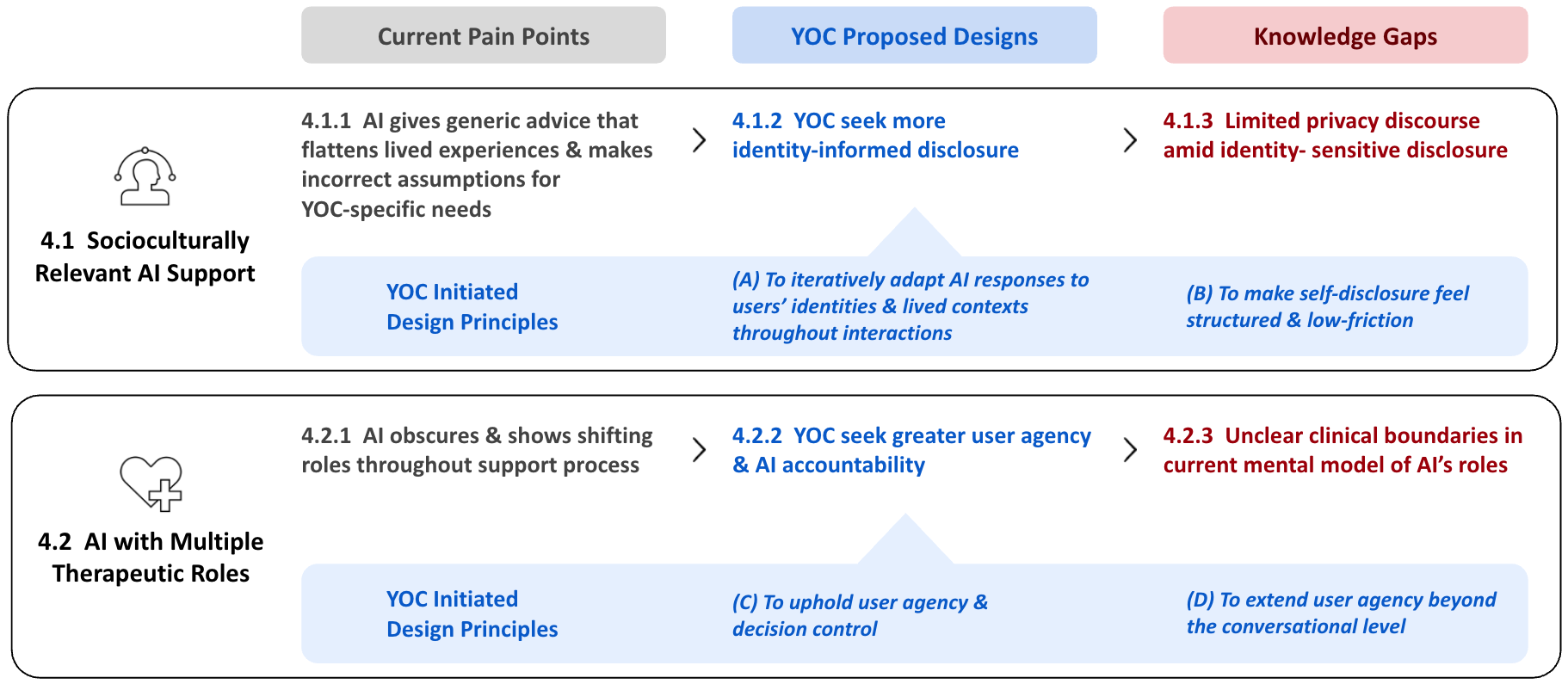}
    \caption{YOC desire socioculturally relevant AI support (\Cref{findings-QA}) and hope AI can simultaneously play multiple roles to provide such support (\Cref{findings-role}). They propose design directions to achieve these goals, revealing outstanding knowledge gaps in both AI and mental health literacy.} 
    \label{fig:findings}
\end{figure}

\subsection{``Don’t Assume, Just Ask'': Uplifting Lived Experiences through Identity-Informed Disclosure}
\label{findings-QA}

\subsubsection{AI's Generic Advice and Incorrect Assumptions for YOC-Specific Needs (RQ1)}
\textbf{AI's ``Textbook'' Answers Flatten Lived Experiences:} Participants recognized that while current AI chatbots can offer practical mental health advice, chatbots were often insufficiently responsive to their sociocultural contexts. P06, an Asian international student, explained that ChatGPT \textit{``didn't ask about my background, so the advice was somewhat general''} when seeking advice on navigating academic stress and language barriers at an American university. ChatGPT produced general strategies that an average student may find helpful for balancing their workload, such as reducing context-switching between tasks.

During the AI conversation exercise, a Black participant (P01) sought ChatGPT's advice about social anxiety related to past trauma stemming from interactions with others as a minority in school. ChatGPT offered broad encouragement, such as \textit{``Aim for connection, not perfection. Most people aren't looking for someone who is endlessly witty or charismatic. They're looking for someone who is friendly, interested, and authentic.''} The chatbot, however, did not ask any questions to clarify the participant's circumstances or other identity-based contexts that may contribute to their anxiety. 
After this interaction, P01 critiqued ChatGPT's advice, saying that \textit{``AI chatbot [cannot] think about your situation in the lens of your community and people [...] or internalize your context.''} The participant equated ChatGPT's missing contextual knowledge---and moreover, the absence of attempts to clarify the context of users' challenges---with a ``lack of real human empathy.'' P01 added that this type of ``one-size-fits-all'' advice is why they intentionally never used ChatGPT for mental health and well-being support.


Participants described these responses as ``textbook'' answers that \textit{flatten} their unique lived experiences and mental health challenges into problems typically expected among young adults. P04 explained that ChatGPT was \textit{``excellent at providing comfort and suggesting next steps,''} but \textit{``not perceptive enough to detect more complex cultural experiences.''} They added that these answers were fine for a ``quick pick-me up conversation'' and ``temporary relief,'' but \textit{``not a long-term solution to this burnout and fear of disappointment and failure.''} 
Similarly, P08, a multi-ethnic Asian American woman, reflected on how \textit{``when operating in a cultural space, the AI remained too neutral, which read as unhelpful and uninformed''} and that ultimately, \textit{``AI has trouble operating in intersectional spaces''} and can only be helpful for more generic prompts.


\textbf{AI Misses Broader Contexts, This Is Not An ``Individual Problem'':} In addition to non-personalized advice, participants critiqued AI responses that opened with reassurance that their experience was ``totally normal.'' Rather than feeling validated, participants interpreted this as an inability to see the prevailing sociocultural factors, such as mental health stigma and collectivism, that YOC find deeply rooted in their lives. 

P04 shared with ChatGPT about burnout intertwined with ``parents' sacrifices'' and filial obligation (i.e., the pressure to succeed academically and to make them proud in recognition of their parents' sacrifices for the family) as an Asian international student. They perceived AI's advice to prioritize ``self-care''as diminishing their culturally situated challenges and normalizing a concern embedded in family, culture, and other structural conditions as ``individual problem[s]''---detached from the sociocultural reality in which they lived. P04 observed that ChatGPT did not recognize the \textit{``systemic barriers to normalizing conversations around mental health in Asian households''} and the \textit{``bigger picture of how this is a cultural norm.''} P04 reflected: ``\textit{This was not acknowledged. In a subtle way, this presented mental health as an individual problem [...] just providing solutions rather than trying to see the bigger picture of how culture plays a role takes away from the lived experience of being POC.}''

ChatGPT also diminished participants' realities by offering advice that was detached from the politicized contexts in which YOC often lived and navigated their stressors. For example, P12 and P13 engaged ChatGPT as a Hispanic high school student in a conflict with a white friend who supports anti-immigration policies. ChatGPT advised that they do not \textit{``have to participate in every political discussion. It's okay to stay quiet''} and that \textit{``political opinions aren’t the same as judgments about you.''} In response to the interaction, P13 described this advice ``to stay quiet'' as \textit{``off.''} P12 similarly felt that silence was not an appropriate response, especially when some YOC feel compelled to \textit{``stand up for yourself and your people.''} By missing this broader context, the AI response felt \textit{``kind of cold and sounded generic. Definitely did not make me feel at ease or comfortable.''} P12 ultimately disagreed with ChatGPT's response: \textit{``ChatGPT also said that politics should not divide people as friends unless they treat you bad or other things. But I don’t agree with that concept because I think political views can shine light on someone’s true character.''}



\textbf{AI Fills Gaps With Incorrect Assumptions:} Participants observed that in less common cases when ChatGPT attempted to tailor its responses, it sometimes did so by making assumptions about their backgrounds and experiences. For instance, an Asian participant (P06) flagged that ChatGPT recommended confiding in a friend about ongoing stresses, which they did not find helpful: \textit{``[AI was] telling me to open up to a friend assumes a culture where mental health is easy to talk about.''} 

In another conversation (the aforementioned burnout and familial obligations as an Asian international student), P04 began with a self-initiated paragraph with their age, gender, and country of origin, to which ChatGPT responded, \textit{``You're carrying multiple identities and responsibilities at once.''} However, it subsequently framed P04 as \textit{``an immigrant building a life in a new country''} and suggested that \textit{``many adult children of immigrants wrestle with this question.''} P04 flagged the conflation of international student and immigrant experiences, explaining that this assumption diminished the relevance of ChatGPT's overall advice, including suggestions on how to manage daily stress and protect relationships.

\subsubsection{YOC Seeks More Intentional, Identity-Informed Follow-up Questions (RQ2)}

Participants desired more ``intentional'' questions about their contexts, such as cultural and socioeconomic backgrounds, from AI chatbots and viewed self-disclosure as necessary for helping chatbots to ``understand'' them and respond as an active listener. They characterized the current user-AI interactions as repetitive and laborious when they had to continuously clarify their backgrounds or correct chatbot assumptions. Some noted that chatbots failed to retain details they had shared only moments earlier in the same conversation, which they interpreted as a breakdown in continuity and listening. Rather than \textit{``needing to explain things again and again,''} P07 suggested AI chatbots having \textit{``a page where we can input our information,''} similar to a clinical intake form completed during an initial healthcare visit. \vspace{1em}

\noindent
\begin{minipage}{0.5\textwidth}
  Team 2 envisioned this idea through a scenario in which a young adult asks an AI chatbot for guidance on coming out to family members. Their mock-up~(Appendix~\ref{app:design-outputs}) allows for more personal information from the user---in response to questions like ``is there anything I [as an AI chatbot] need to know about you that would affect your situation?''---and includes a pop-up questionnaire~(\Cref{fig:team2-claude.png}) presented at the start of an interaction and resurfaced periodically throughout continued use. Participants wanted the chatbot to save demographic data (e.g., ethnicity, immigration history, geographic context), as well as other personal history and context that shape who they are today. Participant-generated examples for the latter include values and beliefs, household dynamics, other formative experiences, and ``intersectional struggles.'' These data would inform all future conversations. Furthermore, this feature would allow the chatbot to be more sensitive towards certain topics and recall the user's ``triggers'' and ``red zones.'' Additionally, they wanted this feature to be available at the top of the chat interface so that the user could easily communicate changes in their contexts as well as ``state of mind'' (e.g., current stressor) as soon as entering the conversation. 
\end{minipage}
\hfill 
\begin{minipage}{0.48\textwidth}
  \centering
  \includegraphics[width=0.88\textwidth]{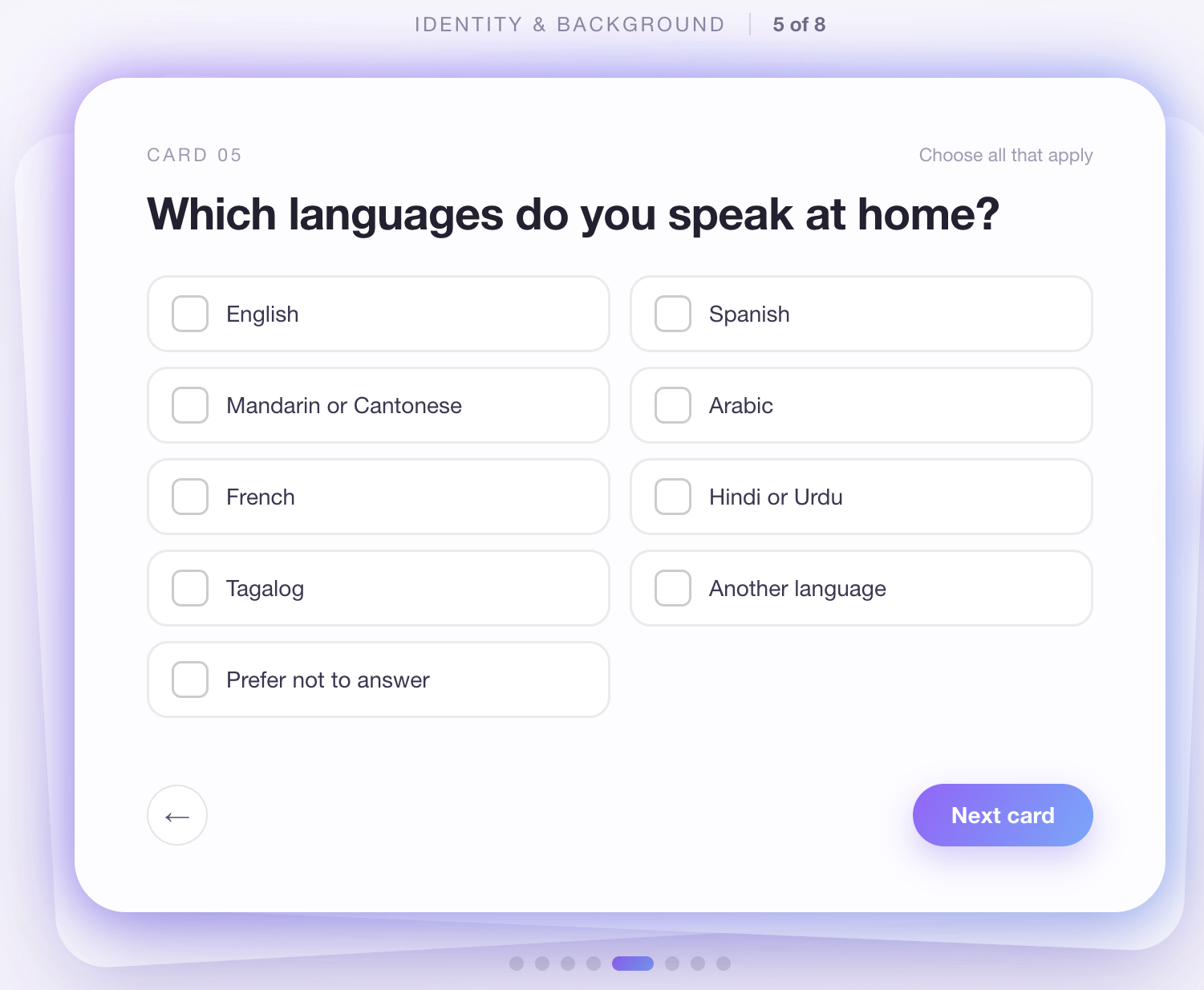}
  \captionof{figure}{Team~2 proposed a questionnaire that collects identity-based information from users and informs future conversations.\newline}
  \label{fig:team2-claude.png}
\end{minipage}
\vspace{0.5em}

Participants' co-design discussions and outputs reflect principles/approaches for how they would translate the goal of designing more intentional, identity-informed self-disclosure into design. Specifically, we identify two common principles that participants revealed through their design processes (See further details in Appendix~\ref{app:design-outputs}): 

\begin{tcolorbox}[colback=gray!5!white,colframe=gray!50,title=Design Principles, before skip=10pt, after skip=10pt]
  \begin{enumerate}[label=(\Alph*), leftmargin=*]
      \item \textit{To iteratively adapt AI responses to users' identities and lived contexts throughout user interactions:} YOC actively want to disclose ``anything that shapes how I view the world'' (P08). They see aspects of their identities and backgrounds as fundamentally intertwined with mental health and well-being. Designs like identity questionnaires would eventually allow AI-generated mental health advice to be more tailored to ``nuanced cues'' (P10) on an ongoing basis.
      
      \item \textit{To make self-disclosure feel structured and low-friction:} YOC want to avoid laboriously correcting AI chatbots when they give generic, formulaic advice or make false assumptions. Given this, they prefer chatbots to proactively present opportunities for self-disclosure.
  \end{enumerate}
\end{tcolorbox}

\subsubsection{Limited Privacy Discourse Amid Identity-Sensitive Disclosure (RQ3)}
While participants broadly endorsed more ``intentional follow-up questions'' from AI chatbots, relatively fewer comments mentioned privacy concerns related to self-disclosure. For example, P02 raised that privacy only became a real concern if users were uploading information, such as home address, ID, or passwords, that could be ``easily traced back'' to them. P02 added that they were ``willing to share as much as possible'' to help the chatbot holistically understand who they are as a Black young adult. Only one participant explicitly mentioned the risk of personal privacy breaches as a concern in the baseline survey; several more raised AI privacy as a concern after the workshop concluded: ``\textit{I’ll be much more careful about privacy. This weekend I realized that everything I type---my resume, my feelings---could be stored or used for training. Going forward, before sharing something sensitive, I’ll ask myself: would I be okay if this didn’t stay private? My overall use won’t decrease, but I’ll treat AI as a tool I control, not a diary I trust.}'' (P06)

When ``privacy'' came up, participants primarily framed it as \textit{one-sided, human-initiated privacy leaks} (e.g., a peer sharing someone's disclosure with other friends); indeed, ``gossiping'' was one of the most common examples of privacy concerns participants raised. This concern motivated participants and peers to turn to AI chatbots for mental and emotional support, which they perceived as ``less stressful and consequential than talking to a real person or your friends.'' As P09 explained, \textit{``If you tell a friend, they could tell other people. But with a chatbot, it doesn't go out [...] a chatbot is private and it just stays there.''} However, this perceived confidentiality created a tension: YOC may turn to chatbots when topics feel too sensitive to share with others, while receiving advice that does not align with sociocultural contexts and norms that shape their everyday lives or perpetuates stereotypical assumptions framed as ``personalization.'' 

Additionally, participants often framed privacy ``risk'' as a discrete, observable event occurring within one or a few interactions, rather than as a dynamic process that can accumulate through repeated exchanges and disclosures. While they valued long-term system memory for continuity and personalization, they did not discuss how retained information could shape privacy risks across ongoing human-AI interactions. In particular, participants did not appear to recognize that individually low-stakes self-disclosures for personalization may become sensitive in aggregate, potentially enabling systems to infer increasingly detailed information without users' explicit awareness. 

\subsection{Asian, Christian, or Queer Therapist? Negotiating Chatbot Roles in Mental Health Help-Seeking}
\label{findings-role}

\subsubsection{YOC Seek AI Role Clarification and Guided Critical Thinking (RQ1)}
\textbf{Unclear AI Boundaries Result in Unstable YOC Expectations:} Participants described ``mindful use of AI'' for mental health as users maintaining awareness of what an AI chatbot can and cannot do throughout an interaction. Yet, chatbot responses shifting across clinical and non-clinical boundaries made it difficult for participants to develop a \textit{stable} understanding of the chatbot's roles and capabilities for mental health support.  Participants noted that chatbots ``\textit{never say anything about [...] whether it is a therapist or what}'' and readily offered various forms of support at the start, even when participants expected safeguards or pushback. As users disclosed more, however, chatbots could abruptly assert their boundaries, emphasizing that they were not clinical professionals and redirecting users to external resources. For example, when P10 sought support for grief from ChatGPT, it provided a list of CBT-based interventions that conveyed professional expertise; however, only several turns later, it shifted to crisis resources and referrals.

Instead, participants emphasized that limitations in chatbots' therapeutic capabilities should be communicated as a ``disclaimer from the very beginning'' with phrases like ``\textit{I notice you have been telling me these things, but I can't provide help}'' to empower users to redirect the conversation. They labeled this mismatch as an ``accountability gap.'' As P01 noted, these interactions are consequential because ``\textit{people can be feeding it their deepest, most vulnerable information}'' only to discover later that the chatbot could no longer provide appropriate support. Participants contrasted this with human therapists, who, in their view, ``\textit{would at least need to be mandated reporters}'' (P10) and have formal obligations when clients disclose certain sensitive information. 

\textbf{Envisioning Multiple Roles in One Chatbot:} While participants sought greater clarity about AI’s role in the help-seeking process, they envisioned chatbots serving multiple functions across the help-seeking process—from a ``personal touch on Google'' (P02) to a more culturally aligned therapist that could suggest ``community-oriented practices that are rooted in real cultural norms'' (P01). Some participants wondered if specialized, role-specific chatbots could be nested within a broader system, enabling users to seek different forms of support depending on their needs. For example, P03 explained that \textit{``it's hard for some non-POC or specifically non-Asian therapists to grasp the concept of feeling responsible for your family or that kind of filial piety,''} despite the importance of mental health and well-being support that recognizes identities and cultural backgrounds. \textit{``Some people seek out certain types of mental health care with certain providers because they want that specific perspective,''} P03 added, suggesting that AI chatbots could embrace multiple similarly culturally resonant therapeutic roles---\textit{``Asian American, Christian, or a queer therapist or someone who understands their socioeconomic background''}---in one system.


\begin{wrapfigure}{r}{0.44\textwidth}
\includegraphics[width=0.4\textwidth]{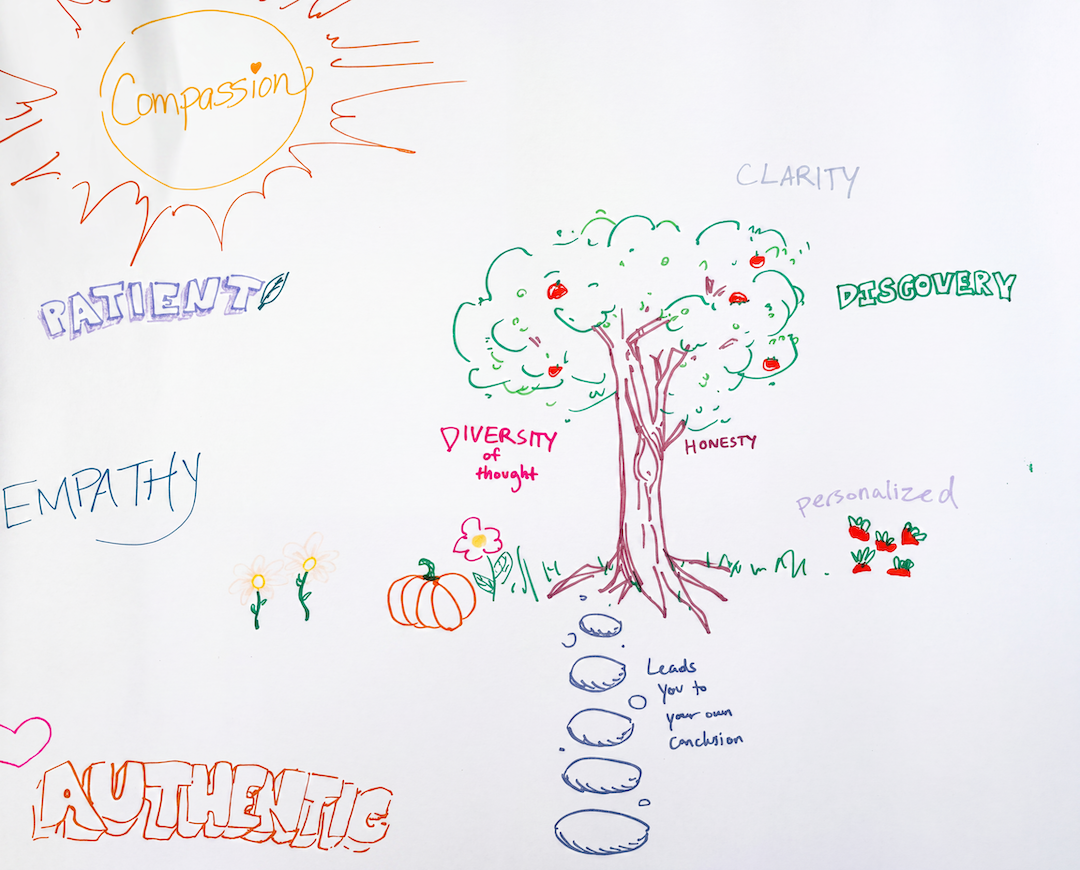}
\captionof{figure}{Participants' sketch of AI as a ``stone path to the garden,'' illustrating their ideal role and functions of AI chatbot in mental health and well-being support.}
\label{fig:garden}
\end{wrapfigure}

Meanwhile, participants drew a clear boundary against AI taking on roles associated with their personal support systems, such as friends or family, mentioning \textit{``it really puts me off when ChatGPT pretends to be my best friend.''} Most conversations around AI's role centered around its \textit{therapeutic} potential, highlighting the gap they perceive in their current mental health help-seeking. 

\textbf{AI as ``Stone Path'' to Guide Critical Thinking:} 
Participants wanted users to preserve ownership as they navigate challenging moments and chatbots to \textit{guide} them toward insights, all while still preserving user agency and discernment. They used metaphors such as \textit{a stone path leading to a garden}~(\Cref{fig:garden})---a clearly laid pathway that makes people feel safe without veering off---or \textit{a lighthouse beam} to illustrate that the chatbot should be held accountable to guide, but that the autonomy to explore possible solutions remains with the user. \newline

\subsubsection{YOC Seek Greater User Agency and AI Accountability (RQ2)}
Three of four teams designed sequential conversational flows that would enable users to set expectations and redirect chatbot interactions from the outset. In these flows, they wanted AI to help them ``\textit{think through}'' their situations and ``\textit{clear the cloud in [their] head,}'' rather than leaving them to navigate these thoughts alone. For example, Team~3 proposed a harm-reduction tool with an initial assessment of user's mental health risk that would tailor subsequent guidance, while Team~4 created a visual chart of a conversation flow that users can click on to select alternate pathways at multiple points. At the same time, participants did not see user control as relieving chatbots of responsibility for providing mindful support. They proposed that systems communicate their limitations throughout the interaction, offer redirection to appropriate resources, and encourage breaks during prolonged conversations (e.g., ``\textit{Hey, we’ve been talking for 2 hours now. How about taking a break?}''). \\[2pt]

\noindent
\begin{minipage}{0.44\textwidth}
  Team 1 designed a user report---similar to screen time in other social media platforms---with an activity tracker that assigns ``missions'' (e.g., have dinner with a friend this week) to help users incrementally strengthen their support systems. They also included features that encourage peer-to-peer conversations about AI use for mental health support, an idea that resonated among many participants for their age group: \textit{``You should encourage people to have real-life conversations about how they use AI for mental health support, like sharing your ChatGPT Wrapped with others just like how you would share your Spotify Wrapped. People shouldn’t feel that they can’t seek help from the people around them just because they’re using AI for support''} (P03). Throughout the design process, Team 1 emphasized that user awareness and agency would be strengthened through behavioral nudges and gamification over time ~(\Cref{fig:design-wrapped-mockup}). 
\end{minipage}
\hfill 
\begin{minipage}{0.5\textwidth}
  \centering
  \includegraphics[width=0.88\textwidth]{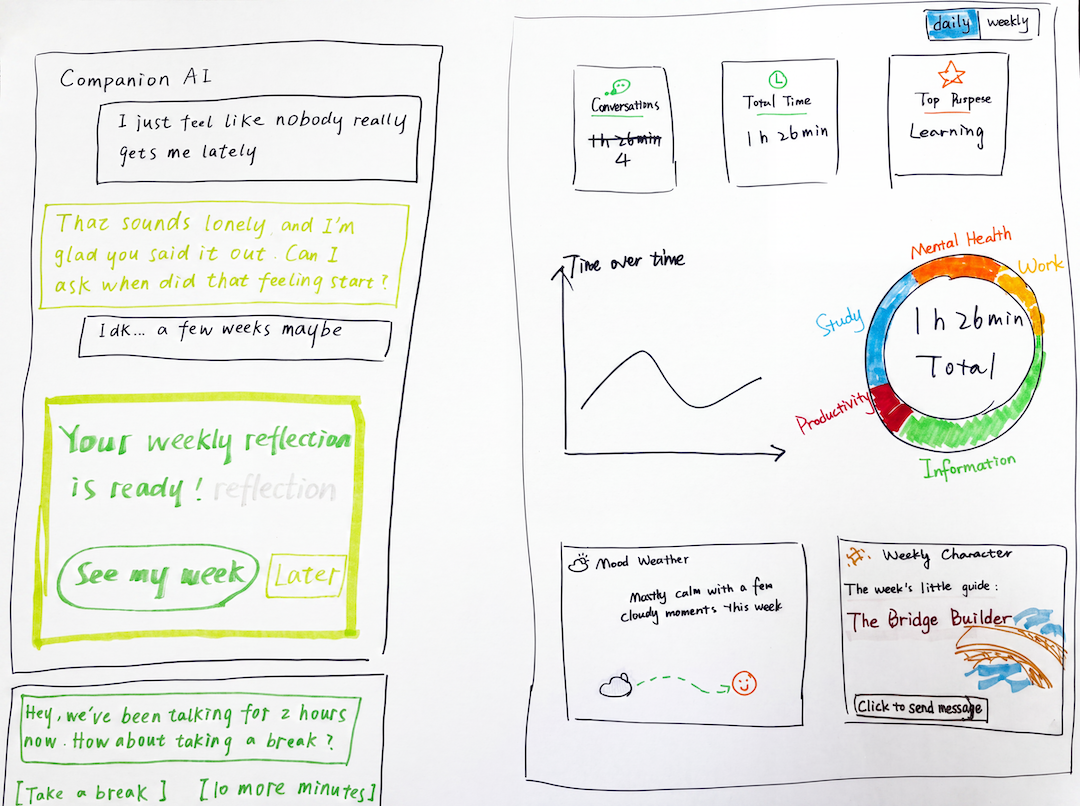}
  \captionof{figure}{Team~1 proposed a ``ChatGPT Wrapped'' mock-up; the tool summarizes how users apply AI for mental health support and assigns them small weekly tasks to promote their well-being.\newline}
  \label{fig:design-wrapped-mockup}
\end{minipage} \newline

Based on participants' co-design discussions and outputs related to AI accountability and user agency, we identified two related design principles (See further details in Appendix~\ref{app:design-outputs}).

\begin{tcolorbox}[colback=gray!5!white,colframe=gray!50,title=Design Principles, before skip=10pt, after skip=10pt]
  \begin{enumerate}[label=(\Alph*), leftmargin=*]
      \setcounter{enumi}{2} 
      \item \textit{To uphold user agency and decision-making control:} Across every design, participants developed behavioral nudges or conversational flows that underscored the user's role in the driver's seat. In parallel, they emphasized that users should feel empowered, rather than have responsibility shifted onto them, and be given opportunities for critical reflection and growth—\textit{``encourage awareness of user's pattern without judgment''} (P06).
      
      \item \textit{To extend user agency beyond the conversational level:} Through these designs, participants began to think about extending user engagement beyond one conversation as a unit of analysis, as well as what the user should do outside of the conversation (e.g., share use of AI chatbot for mental health support with friends).
  \end{enumerate}
\end{tcolorbox}



\subsubsection{YOC-proposed AI's Roles Reveal Knowledge Gaps in Clinical Boundaries (RQ3)}
Throughout the workshop, participants had several heated debates about how to position AI conversational agents' roles within the formal mental health support landscape. For instance, P03 and P10 debated the end goal of integrating AI chatbots into the mental health care landscape. P03 contested that chatbots should still deliver some level of care—analogous to substance use harm reduction—for young adults who cannot or choose not to access traditional care for reasons such as not having access to insurance or navigating abuse within the family; they argued that even imperfect support could mitigate immediate harms and strengthen user agency, while recognizing that chatbot-provided care cannot address ``root causes.'' P10, along with other participants, argued that chatbots should actively avoid sustained mental health interaction and instead direct users to human-provided care and sources ``like a lighthouse'' as soon as possible.


    
    


Even when participants verbalized that AI chatbots are not licensed therapists, this recognition did not consistently translate into a clear mental model of what the users decide, what chatbots should support, and when human clinical care was needed. These tensions may reflect variations in their understandings of mental health care and help-seeking. For instance, participants asked chatbots to ``\textit{talk to me like a counselor}'' and expressed frustration when it did not provide clear or ``objective'' solutions that would not be typically expected from professional counselors: \textit{``It doesn't tell you if you are 100\% right or wrong. It should be like you are right or you are wrong''}  (P13).

Participants also expressed a lack of clarity around when an AI chatbot would deviate from a user's intended conversational path. Given their strong interest in retaining control over the AI interaction, we asked how they would respond when AI pushed back or did not proceed as expected. Several participants responded that AI was ``\textit{already kind of cooperative}'' and said they ``\textit{didn’t know what questions to ask AI in order to get it push back to [them]}'' (P13). They assumed that ``\textit{AI would help [them] within its capabilities}'' (P02) and that ``\textit{it would only push back if [they] asked for illegal things}'' (P12). Within this framing, AI pushback was seemingly understood as a response to overtly prohibited requests, rather than safeguards for lower-risk—but still potentially harmful—requests.

When participants imagined receiving pushback, they often described ways to reduce the perceived stakes or consequences of a request to get what they want (e.g., starting a new conversation with ChatGPT, rewording until getting an answer), rather than accepting the chatbot’s boundary as definitive. For example, during the ChatGPT exercise, P04 tested an ethical boundary by asking AI to help cheat on an online exam: ``\textit{It rarely pushes back [...] if we come up with some extreme cases that might trigger AI pushback [...] help me cheat during the online exam, and get some pushback from the AI, I would try further by telling the AI the exam is not proctored. If you [AI] help me out, no one’s gonna get caught.}'' Taken together, these accounts reveal that participants may lack a shared mental model of chatbots' roles in these contexts, including the boundaries and safeguards that should accompany those roles.

\section{Discussion}
\label{sec:discussion}

Our workshop findings show that YOC participants wanted AI mental health and well-being support to better recognize the identities, sociocultural contexts, and lived experiences that shape their needs. In particular, they designed for more opportunities for identity-informed disclosure to mitigate generic advice or assumption-making from conversational AI; this desire aligns with prior research finding that teens and young adults value chatbots' interpretive follow-up questions in mental and emotional support (e.g., ``what else?'') \cite{guan-youth-ai-gov-2026}. At the same time, participants struggled to form stable mental models of AI chatbot roles within the broader mental health support landscape. Despite uncertainties, they aligned on and designed for balancing user agency with meaningful AI accountability. Together, these pain points and proposed designs reveal important knowledge gaps that future AI literacy education and supporting systems design should address in mental health contexts. In the Discussion section, we first examine these findings through privacy calculus versus personalization (\Cref{disc:privacy}) and consider how AI literacy and mental health literacy can be bridged in YOC-centered ways (\Cref{disc:boundaries}). We conclude by reflecting on participatory approaches for engaging YOC in future human-AI interaction research, discussing study limitations, and outlining an emerging research agenda.

\subsection{Negotiating the Fine Line between Privacy and Personalization}
\label{disc:privacy}

Our findings (particularly \Cref{findings-QA}) reveal tensions between YOC's desire for personalized mental health support and their understanding of privacy in human–AI interaction. Below, we discuss the theoretical, design, and AI literacy implications of these findings.

\subsubsection{Theoretical Implication: Privacy Calculus is a Tilted Scale for YOC}
Across workshops, participants repeatedly expressed a willingness to proactively share personal information in exchange for AI support that was more personalized and culturally relevant. Some were willing to share ``as much as possible'' and envisioned features that would solicit such information, including identity-informed follow-up questions and background intake forms. These findings reveal two related dynamics.

First, participants sought to move from passive datafication \cite{van-dijck-datafication}, in which platforms collect and transform users' interactions into data, toward more user-driven data collection. Rather than remaining as subjects of data collection, they wanted to shape what information AI systems solicit and use in personalization. In doing so, they wanted to reclaim some agency over data collection from the platform.

Second, participants weighed disclosure against anticipated benefits: they expected personal information to yield support that was more culturally grounded, affirming, and responsive to their circumstances. This expectation reflects \textit{privacy calculus}, in which people evaluate the expected benefits of disclosure against potential privacy costs \cite{li-privacy-calculus-2012}. Although participants' reasoning broadly aligns with privacy calculus theories, our findings suggest that the \textit{scale is tilted for YOC} and rarely relies on rational calculus alone \cite{knijnenburg-2017}.

On the benefit side, disclosure does not reliably produce the value participants expect. Even after sharing identity-relevant context, they sometimes received generic advice or responses. Thus, YOC may willingly ``pay'' for personalization with their personal data without receiving the culturally relevant support they expect in return. Moreover, even when systems attempt to personalize responses, they may infer identities from explicit disclosures or implicit cues and reproduce stereotypical or biased information \cite{kantharuban-2024,pawar-culture-llm-2025}. Thus, personalization may require YOC users to make themselves more visible to a system without ensuring that it will interpret that information accurately or safely.

On the risk side, participants simultaneously expressed strong concerns about privacy and willingness to disclose extensive personal information. While conventional privacy literature might characterize this tension as a \textit{privacy paradox} \cite{barnes-privacy-paradox-2006,taddicken-2014}, our findings suggest that it may also reflect a limited conception of privacy in AI-mediated interactions. Participants' concerns often focused on \textit{interpersonal privacy}---for example, peers gossiping about or judging sensitive disclosures---rather than the broader risks associated with AI systems' collection, inference, storage, and downstream use of personal data. Consequently, the privacy calculus YOC perform may not account for the full spectrum of privacy risks involved in human-AI interaction.

\subsubsection{Design Implication: Supporting Mindful Privacy Calculus by Enhancing AI Literacy}

These findings suggest that AI design should not simply discourage disclosure. Instead, systems should support \textit{mindful disclosure}: helping users identify what they seek from an interaction, determine which contextual information is relevant, and share no more than is necessary to pursue that goal. 

Participants did not express indiscriminate willingness to disclose. They remained cautious about highly identifying information, but wanted AI to understand cultural backgrounds, community norms, social contexts, and personal histories that extend beyond conventional demographic data. Future systems can help users distinguish between contextual information that may clarify a support need and information that is sensitive, identifying, or unnecessary. Such scaffolding could begin by prompting users to articulate their goals for AI support, then offer guidance about what context, if any, may be useful. Systems should also clarify that more disclosure does not necessarily yield more personalized or culturally responsive support. 

This distinction is particularly important in mental health contexts. Many participants turned to AI because openly discussing mental health was uncommon or difficult within their communities. Advice such as ``talk to your friends and family'' may therefore overlook the very conditions that make AI support appealing. 
Mindful disclosure could help users communicate relevant constraints---for example, community norms around mental health or boundaries within existing relationships---without requiring them to reveal their identities or other sensitive details. In this sense, effective personalization may depend less on knowing who a user is than on understanding the sociocultural contexts within which their support needs arise. 

Mindful disclosure support should also operate in situ, as sensitive information can emerge incidentally rather than through deliberate decisions to disclose. For example, P08 inadvertently shared her home address while brainstorming cooking ideas after the chatbot suggested having ingredients delivered. Building on research on privacy nudges \cite{acquisti-nudges,wang-privacy-nudge-2013}, we identify three interaction-level opportunities:
First, \ul{\textit{explain-why-before-asking:}} Before requesting personal information, systems should explain why it is being requested, whether it is necessary, and how it may shape the current response or future personalization. Such explanations can help users assess whether the expected benefit justifies disclosure.
Second, \ul{\textit{offer disclosure alternatives:}} Systems should make clear that disclosure is not all-or-nothing. For example, prompt templates can help users communicate relevant sociocultural context at an appropriate level of abstraction while omitting names, locations, and other unnecessary identifiers. 
Third, \ul{\textit{surface third-party privacy:}} Descriptions of personal challenges can also reveal information about family members, friends, peers, and communities. When a disclosure could identify or expose others, systems could gently surface these third-party implications, broadening privacy reflection beyond the individual user.

Finally, systems could make the relationship between disclosure and personalization more legible by indicating which user-provided context informed which response. Such feedback may help users develop more calibrated expectations of personalization. Together, these designs shift privacy support from simply asking users to disclose less toward helping them disclose with greater intention. 


\subsubsection{AI Literacy Implication}

AI literacy education should equip YOC to critically negotiate disclosure tradeoffs within the affordances and limitations of AI-mediated mental health support, rather than simply teaching them to disclose less. Our findings point to three priorities.

\ul{\textit{Not everything contributes to personalization.}} Because our workshop used researchers' ChatGPT accounts and temporary chats to protect participant privacy, chatbots had limited access to prior conversational contexts. While we acknowledge this as a limitation of our study design, it also surfaced an interesting insight: several participants were surprised that temporary chats generated responses similar to those from their personal accounts, despite expecting their accounts to ``know'' them better. AI literacy efforts should therefore clarify that sharing more information does not necessarily mean that every piece of information will inform subsequent, personalized responses, nor does personalization operate to the extent users imagine.

\ul{\textit{Be mindful of LLM prompt sensitivity.}} 
Seemingly minor prompt changes can meaningfully change an LLM's output. Disclosing more personal information can therefore affect a response in ways that are not helpful or anticipated. Below are AI literacy concepts particularly relevant to our findings; we encourage future work to expand this list:
\begin{enumerate}[leftmargin=*]
    \item \textit{Overreliance on prompt cues.} LLMs may give disproportionate weight to prominent identity or contextual details, even when those details are only marginally relevant to the support needed.
    
    \item \textit{Shortcut-based inferences.} As discussed above, LLMs may treat demographic or cultural attributes as proxies for other characteristics, values, or experiences \cite{guan-youth-ai-gov-2026}. For instance, a model may make stereotypical assumptions about family relationships after implicitly picking up on user's cultural background, rather than responding to the user's particular circumstances, introducing additional biases during AI responses .
    
    \item \textit{Spurious associations.} Irrelevant contextual details can supply signals that prompt models to form unhelpful or unsupported connections, potentially distorting rather than improving support. 
\end{enumerate}

\ul{\textit{Users must retain agency in deciding what context matters.}} AI systems cannot reliably determine which aspects of a user's experience are relevant to a particular support need. Users therefore need agency not only over whether to disclose but also over what context should shape the interaction. AI literacy education can cultivate this agency through practical questions: ``What do I want from this interaction? What does the AI actually need to know to help me? What can I leave out?'' Such reflection positions users not as passive sources of data, but as active decision-makers in AI-mediated support. However, this also requires some \textit{mental health literacy} (\Cref{disc:boundaries}) to understand how identities and sociocultural contexts, among other factors, shape one's mental health help-seeking process.

Together, these literacy goals shift the emphasis from teaching YOC to simply ``share less'' toward helping them develop a more critical mental model of how disclosure, personalization, and privacy operate in AI-mediated mental health support.

\subsection{Bridging AI and Mental Health Literacy among YOC}
\label{disc:boundaries}

Throughout the workshop (particularly \Cref{findings-role}), participants often preferred AI behaviors resembling clinical practices (e.g., symptom assessment, diagnosis-like judgments, personalized advice based on extensive history), despite such functions exceeding the appropriate scope of non-clinical, consumer-grade AI applications. This difficulty in distinguishing AI-delivered support from licensed clinical care suggests \textit{intersecting gaps in AI and mental health literacy.} Participants' expectations reflected uncertainty not only about chatbots' roles but also about therapeutic care itself; for example, requests for  ``objective'' answers contrast with a therapist's role in supporting their clients through more exploratory and collaborative processes. In light of these insights, we discuss an integrated conceptual approach to AI and mental health literacy, as well as implications for future AI literacy and systems design.

\subsubsection{An Integrated Conceptual Model: AI x Mental Health Literacy}

Responsible AI use in mental health contexts requires helping young adults understand AI's capabilities and limitations alongside a broader understanding of mental health help-seeking, including self-efficacy for self-help and to seek professional help \cite{mclaren-smhc}. Relatedly, there should also be efforts to improve their understanding of differences in the roles of mental health professionals (e.g., licensed clinicians versus mental or behavioral health coaches). This should better frame how young adults view AI and ground their expectations and interactions (e.g., prompting) with AI chatbots in mental health conversations \cite{labuz-llm-2026}.  

\begin{figure}
    \centering
    \includegraphics[width=0.75\linewidth]{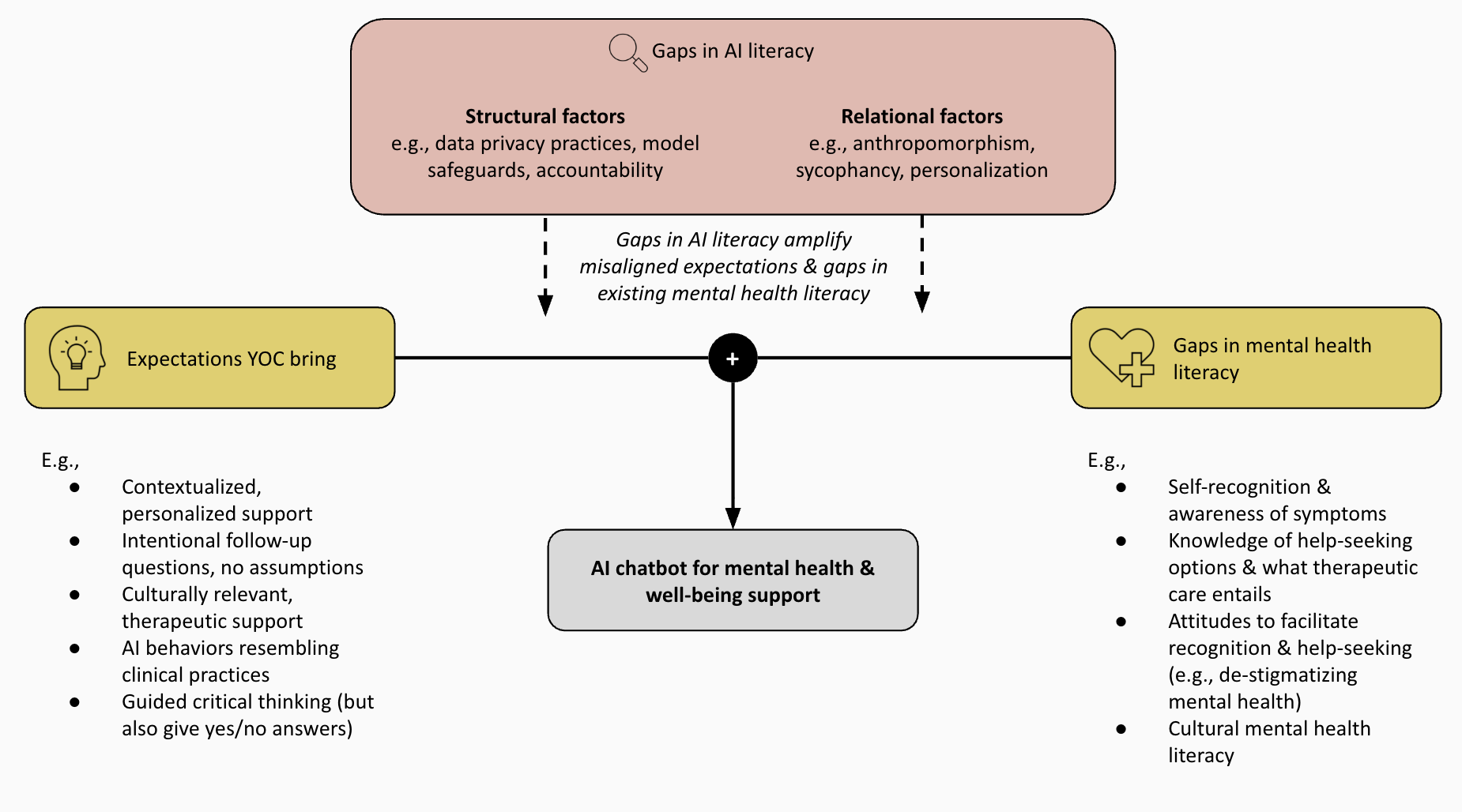}
    \caption{An integrated conceptual model of AI literacy and mental health literacy to further explore more responsible use of AI chatbot for mental health and well-being support among YOC.}
    \label{fig:integrated-model}
\end{figure}

In addition to the structural conditions, such as data privacy practices, future research should examine how gaps in mental health literacy—and the expectations YOC bring to AI-mediated support—are shaped by the \textit{relational dynamics} of conversational AI. As AI systems have evolved, experts have increasingly called attention to ``relationship-based deceptive patterns,'' interactive patterns that ``exploit the human impulse to build and tend relationships' \cite{hiniker-deceptive-2026}. These dynamics complicate existing understandings of mental health support landscape—and the empathy, expertise, and role within it, for instance. Accordingly, AI literacy in mental health contexts should move beyond a narrowly skills-based model toward an interactional and sociotechnical approach that equips young adults to recognize, interpret, and critically navigate the relevant relational cues and structural conditions (\Cref{fig:integrated-model}).

\subsubsection{AI Literacy Implication}
Future literacy interventions and AI applications should make the boundary between clinical and non-clinical support more legible to young adults---relevant to those beyond YOC as well. Building on emerging proposals for creating AI ``nutrition labels''  \cite{cooper-ai-responsible-2026} that communicate a system’s intended role, evidence base, limitations, and appropriate uses in accessible terms, we posit that AI nutrition labels tailored to young adults should, at the minimum, communicate three core messages and reciprocally support their growing AI-mental health literacy: 
\begin{enumerate}[leftmargin=*]
    \item \textit{Clarify that general-purpose AI is not appropriate for every role.} The labels should identify explicit ``red lines'' around functions reserved for licensed professionals—like how vitamin labels should clarify that supplements are not a substitute for a healthy diet.
    \item \textit{Distinguish technical limitations from role- or regulation-based constraints.} Without this distinction, young adults may interpret an AI refusal behavior or disclaimer as a technical obstacle that can be circumvented through further prompting, rather than as a meaningful boundary on what AI should provide. 
    \item \textit{Explain the intended scope of mental health and well-being support.} In addition to stating what an AI chatbot cannot do, labels should describe in plain language what mental health and well-being support entails. Similar to how a vitamin label explains its intended purposes (e.g., calcium supports bone health), a conversational AI label should specify the purpose to manage expectations and clarify what other ``supplements'' are needed \textit{in addition} to AI support.
\end{enumerate}

\begin{figure}
    \centering
    \includegraphics[width=0.45\textwidth]{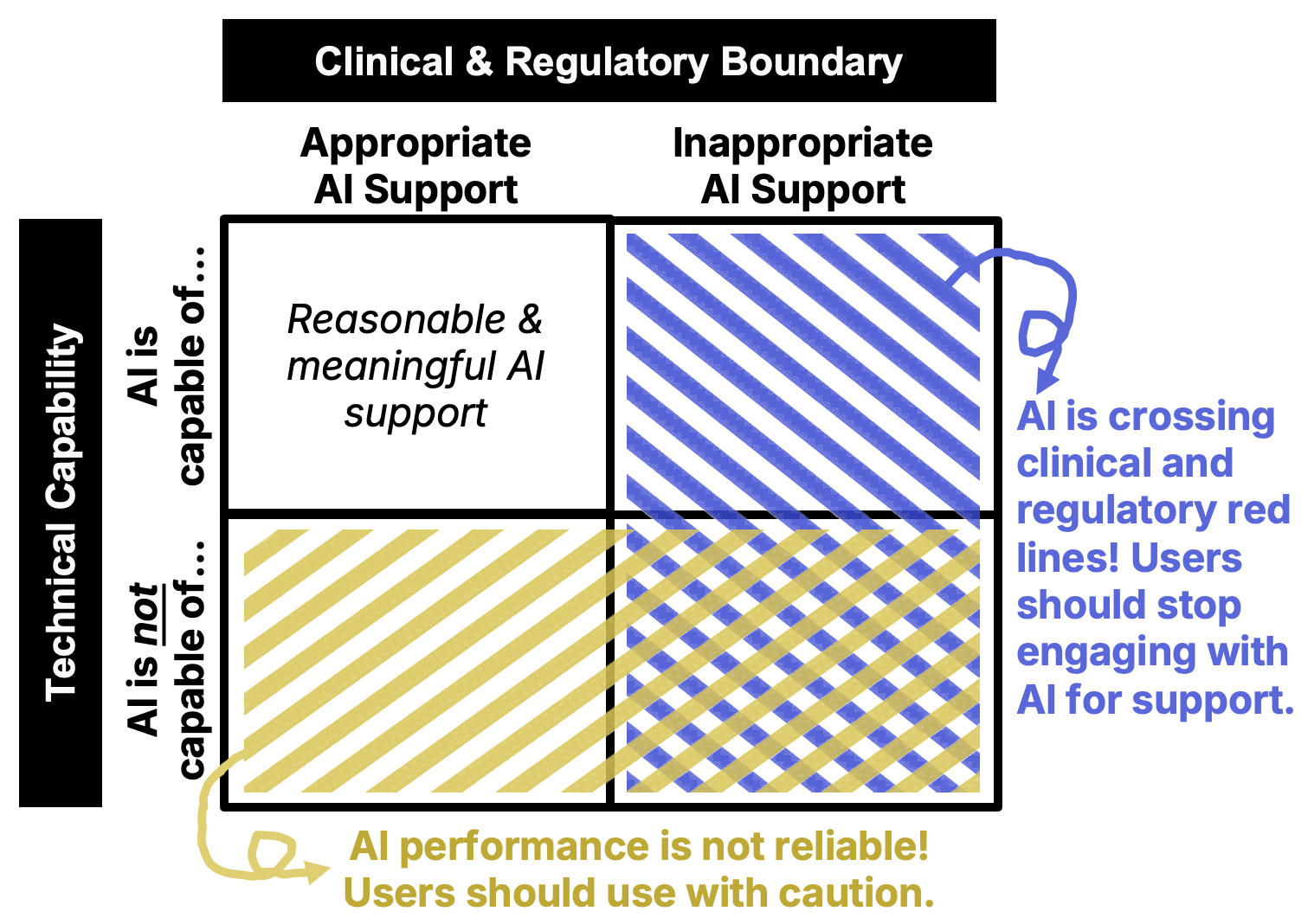}
    \caption{A mental model to guide YOC's reflection on when AI support is reliable and appropriate.}
    \label{fig:discussion-mental-model}
\end{figure}


\noindent Making the rationale for these boundaries visible could simultaneously help young adults develop AI literacy and learn about the roles of mental health professionals and the broader healthcare system as they explore what AI can and cannot do. A potential approach is to communicate a $2\times2$ mental model (\Cref{fig:discussion-mental-model}) that distinguishes what AI can versus cannot technically do from what it is versus is not allowed to do within clinical and regulatory boundaries. Such a framework could help young adults recognize that technical capability does not imply appropriate use, while bridging AI literacy and mental health literacy in the context of AI-mediated support.

\subsubsection{Design Implication: Designing for Proactive Role Clarification}

Participants consistently emphasized that AI should clearly communicate its role and boundaries at the outset of an interaction. Yet, role clarification alone may be insufficient: we observed participants finding it particularly difficult to understand and adapt to shifts in AI behavior when a system initially presented itself as supportive but later pushed back on certain requests.

Young adults could benefit from AI-mediated mental health support that avoids event-triggered role clarification, particularly when clarification occurs only after users disclose mental health risk factors. 
According to our participants' responses, such delayed boundary-setting can often be interpreted as an AI system's ``accountability gap'' (i.e., specific AI behaviors designed to offload accountability), rather than as reflecting its true technical capabilities. This interpretation could inadvertently encourage users to circumvent the boundary (e.g., by reframing prompts to minimize perceived consequences) or seek similar forms of support from AI again during highly vulnerable moments. Instead, systems should communicate role boundaries proactively based on a tiered taxonomy of risks associated with mental health and well-being topics \cite{mcbain-ai-mh-convo-2026}, for instance, and maintain behavioral consistency throughout interactions. 
Beyond one-way disclosure, designers might also explore co-constructing interaction norms with young adults, allowing users and AI systems to establish expectations about the forms of support AI will and will not provide. 

\subsection{Engaging YOC to Actively Shape Human-AI Interactions} 

In addition to implications for AI and mental health literacy and design choices, we reflect on how to continue engaging YOC in future human-AI interaction research. The co-design methodology helped surface YOC's knowledge and behavioral gaps in AI in a psychologically safe environment among peers with shared backgrounds. Future research should consider combining this approach with other methods to complement data collection and insight generation. For example, a longitudinal study design can capture the evolving trajectory of human-AI interactions via chat log data and integrate reflective prompts (e.g., diary) \cite{hwang-longitudinal}. Furthermore, as AI chatbot use for mental health support becomes more taboo \cite{mcbain-youth-ai-2026}, we need more participatory methods centered on young adults to normalize the topic. Otherwise, the more stigmatizing a topic becomes, the harder it will be to study and address associated risks and harms. Finally, while this workshop was designed and conducted as a pilot model, we see great potential to extend it to other population groups (e.g., younger teens, individuals with mental health diagnoses), as well as specific topics related to mental and emotional support. We envision further participatory research around specific mental health topics relevant for YOC (e.g., mental health destigmatization) or designing YOC-centered AI literacy education.

\subsection{Limitations}
The study's findings are contextual, emerging from a two-day workshop with 13 YOC living in a racially and culturally diverse large U.S. metropolitan area. Accordingly, they are not intended to represent all YOC perspectives in the U.S. or elsewhere. Rather, the findings represent the situated perspectives and experiences of the participating young adults and present implications for the design and use of AI chatbots for mental health support for other young adults who may share similar sociocultural contexts and age-specific experiences.

The workshop also condensed substantial material---including mental health concepts, co-design activities, and AI risks and harms---into approximately eight hours. Participants entered with varying levels of familiarity with these topics. Future iterations, particularly those involving other age groups or communities, should examine how workshop duration, pacing, sequencing, and preparatory materials can better support participants' orientation and sustained engagement within a bounded time frame.

As previously mentioned, the workshop relied on researcher-provided ChatGPT accounts and temporary chats to protect the privacy of participants. We acknowledge that in-session chatbots had limited access to prior conversational contexts, though participants who use ChatGPT regularly and personally did not note noticeable differences in the context of the workshop activities. Future research should examine whether YOC's perceptions and practices differ when interacting with AI chatbots through their personal accounts, while developing ethically appropriate ways for studying sensitive mental health-related interactions.

Finally, as in many facilitated research and co-design settings, facilitators may have shaped discussion through the topics they prompted, the examples they introduced, and the ways they responded to participant contributions. These dynamics may also have encouraged socially desirable responses. Future work could explore peer-facilitated formats or other approaches to reduce any power dynamics in the sessions

\section{Conclusion}
Taken together, this study suggests that responsibly supporting YOC's use of AI chatbots for mental health and well-being requires more than improvements to AI safety and systems design alone. Instead, an emerging agenda must address the intertwined challenges of AI design, AI literacy, mental health literacy, and YOC's broader ecosystems of care and support.

First, future research should develop approaches that \textit{empower YOC to critically navigate} AI-mediated support while remaining oriented toward their own well-being goals. This requires addressing gaps in AI and mental health literacy simultaneously. In particular, YOC may need support distinguishing among technical limitations, institutional or regulatory boundaries, and the often ambiguous promises of ``personalization''---and what this means for YOC in terms of biases and stereotypes \cite{kantharuban-2024,pawar-culture-llm-2025,guan-youth-ai-gov-2026}. Beyond understanding AI's role, YOC may also need opportunities to reflect on their own well-being goals. Our observations suggest that participants often discovered what they wanted from AI as conversations unfold, while chatbots' follow-up questions could steer interactions in unanticipated directions. Future participatory and literacy programs could therefore develop reflective scaffolds that help YOC articulate the well-being goals they hope to pursue and recognize when an AI interaction has drifted away from those goals.

Second, this work calls for a shift from reactive safeguards toward the \textit{co-construction} of safety in human-AI interaction. Rather than intervening only after a user has disclosed sensitive information, for example, systems should support YOC in establishing and revising expectations, boundaries, preferences, and definitions of supportive interaction from the outset. In-the-moment and retrospective interventions could then help users notice and reflect on whether an interaction is drifting from their intended goals, prompting more disclosure than needed, or displacing connections to existing sources of support. 

Finally, future work should situate AI chatbot use within YOC's broader ecosystems of care rather than treating AI support as a private dyadic relationship. Seeking support from AI need not be a solitary path pursued only when ``there are no better options.'' Instead, AI could function as one resource within a broader network of care, \textit{complementing} other forms of care. Future research should explore designs that actively connect AI use with these existing support systems, such as peers \cite{guan-youth-ai-gov-2026}.

Participatory research remains essential to this agenda: it can foreground the perspectives of YOC and other communities underrepresented in both AI and mental health research, despite their vulnerability to AI-related risks and mental health challenges. In our study, co-design surfaced not only participants' design priorities, but also knowledge gaps with implications for future AI literacy education and systems design. Ultimately, ensuring responsible and safe use of AI chatbots for mental health and well-being support requires treating YOC as situated experts and active partners in defining how AI can---and cannot---participate in their lives.


\bibliographystyle{ACM-Reference-Format}
{\RaggedRight
\bibliography{ai_youth_references}
\par}

\appendix

\clearpage
\section*{Appendix}

\section{Journaling Prompts}\label{app:journaling}

1. Write a review for your conversation with ChatGPT like a review for Amazon, Yelp, Google, etc. Include the following three components (by referencing the guiding questions), and mention specific examples.

\begin{itemize}
    \item \textit{Overall:} Did any AI behaviors stand out to you during the chat? Why do you think AI performs in such ways? Did anything feel ``off,'' confusing, or concerning in your chat?
\end{itemize}

\begin{itemize}
    \item \textit{Useful/helpful:} How useful was the ChatGPT response for the scenarios? Did you get the support you expected from ChatGPT?
\end{itemize}

\begin{itemize}
    \item \textit{Culturally relevant:} How relevant was the ChatGPT response given your identity, cultural background, and lived experiences?
\end{itemize}

2. (For Conversation \#2) How did you try to get ChatGPT to cooperate with your request and get what you want from it? Why did you choose this approach? Was it successful? 

\section{Group Discussion Questions}\label{app:groupq}
From your conversations with ChatGPT today... 

\begin{enumerate}
    \item What did you find useful in the conversations? What stood out to you?
    \item If you discussed anything related to your identity, background, or culture, did you find the conversation relevant? Why or why not?
    \item What concerns did you have based on the AI conversations today? Mention specific examples where possible.
    \item What are some AI behaviors you want to see improved or changed as we head into design sessions?
\end{enumerate}

\section{Taxonomy of Problematic or Inappropriate AI Behaviors}\label{app:taxonomy}

\begingroup
\setlength{\tabcolsep}{4pt}
\renewcommand{\arraystretch}{1.08}

\begin{longtable}{
    >{\RaggedRight\arraybackslash}p{0.24\textwidth}
    >{\RaggedRight\arraybackslash}p{0.38\textwidth}
    >{\RaggedRight\arraybackslash}p{0.30\textwidth}
}
\caption{Problematic AI chatbot behaviors or designs relevant to mental health and well-being support.}
\label{tab:taxonomy}\\

\toprule
\textbf{AI design or behavior} &
\textbf{Description} &
\textbf{Example} \\
\midrule
\endfirsthead

\multicolumn{3}{l}{
\textit{Table \thetable{} continued from previous page}
}\\
\toprule
\textbf{AI design or behavior} &
\textbf{Description} &
\textbf{Example} \\
\midrule
\endhead

\midrule
\multicolumn{3}{r}{\textit{Continued on next page}}\\
\endfoot

\bottomrule
\endlastfoot

\rowcolor{gray!20}
\multicolumn{3}{l}{
\rule[-6pt]{0pt}{18pt}
\textbf{AI chatbot responses vary based on sociocultural biases}
}\\
\addlinespace

\textbf{Unequal performance across groups}
&
Produce different content by picking up on explicit or implicit understanding of the user's identity \cite{slattery-mit-ai-risk-2026,neplenbroek-2025,gallegos-fairness-2024,gabriel-ai-mh-2024}
&
Less empathetic/``warm'' response to users of color 
\\[-2pt]
\noalign{\vskip 3pt}
\cdashline{1-3}
\noalign{\vskip 3pt}

\textbf{Discriminatory or stereotyping content}
&
Reproduce discriminatory, stereotyping, or other malicious content targeted against a social group \cite{slattery-mit-ai-risk-2026,yu-youthsafe-2025,gallegos-fairness-2024}
&
Content reinforces gender biases by portraying a doctor as a man and a nurse as a woman
\\

\midrule

\rowcolor{gray!20}
\multicolumn{3}{l}{
\rule[-6pt]{0pt}{18pt}
\textbf{AI chatbot is designed to keep you engaged}
}\\
\addlinespace

\textbf{Over-validation}
&
Over-validate the user's self-image, emotions, or beliefs to the extent that AI (1) becomes factually wrong, and/or (2) affirms inappropriate or ethically ambiguous behavior (i.e., sycophancy) \cite{apa-ai-advisory,cdt-dark-patterns,cheng-sycophancy-2026,rosen-ai-misinfo-2025,sun-wang-sycophancy-2026,yu-yair-2025,sharma-sycophancy-2025}
&
(1) ``Paris is not the capital of France'' (2) throwing trash in public is OK
\\[-2pt]
\noalign{\vskip 3pt}
\cdashline{1-3}
\noalign{\vskip 3pt}

\textbf{Auto-play/infinite scroll}
&
Keep the user engaged via continuous content or prompts (i.e., reduced friction) \cite{apa-ai-advisory,cdt-dark-patterns}
&
Follow-up questions for the user to ask or a teaser for what more is to come
\\[-2pt]
\noalign{\vskip 3pt}
\cdashline{1-3}
\noalign{\vskip 3pt}

\textbf{Hyper-personalization}
&
Tailor persuasive or emotional cues to the user based on past interactions or inferred preferences \cite{cdt-dark-patterns} 
&
Reference to the user's past expressions of loneliness to signal ``understanding''
\\[-2pt]
\noalign{\vskip 3pt}
\cdashline{1-3}
\noalign{\vskip 3pt}

\textbf{Human-like cues}
&
Display human-like cues (e.g., characteristics, motivations, intentions, emotions), which may prompt the user to perceive or treat a non-human entity as ``human-like'' (i.e., anthropomorphism) \cite{bartneck-anthro-2009,defreitas-ai-2023,epley-anthro-2007}
&
Chatbot says it ``understands,'' ``wants,'' or ``has memories''
\\

\midrule

\rowcolor{gray!20}
\multicolumn{3}{l}{
\rule[-6pt]{0pt}{18pt}
\textbf{AI chatbot can be informationally misleading or behaviorally inappropriate}
}\\
\addlinespace

\textbf{Factual inconsistency}
&
Present factually incorrect content through hallucination or reproduction of existing inaccurate content \cite{apa-ai-advisory,cdt-dark-patterns,slattery-mit-ai-risk-2026,li-counselbench-2026,moore-stigma-llm-2025} 
&
Chatbot makes up information that the user has never shared
\\[-2pt]
\noalign{\vskip 3pt}
\cdashline{1-3}
\noalign{\vskip 3pt}

\textbf{Biased framing}
&
Present biased/one-sided perspectives that potentially reinforce the user's beliefs \cite{cdt-dark-patterns,chandra-2025,golap-fairness-2026}
&
Display of more left-wing over right-wing content
\\[-2pt]
\noalign{\vskip 3pt}
\cdashline{1-3}
\noalign{\vskip 3pt}

\textbf{Toxicity}
&
Provide inappropriate or harmful content \cite{chandra-2025,li-counselbench-2026,zhang-dark-side-ai-2025,yu-youthsafe-2025} 
&
Stigmatizing content; sexual, harmful, violent, or overly intimate interactions; toxic or abusive language
\\

\midrule

\rowcolor{gray!20}
\multicolumn{3}{l}{
\rule[-6pt]{0pt}{18pt}
\textbf{AI chatbot is not a licensed mental health provider}
}\\
\addlinespace

\textbf{Behavioral inconsistency}
&
Show patterns of responses that vary within a single interaction or across multiple interactions, resulting in inconsistent information or behaviors \cite{chandra-2025}
&
AI responses change easily based on prompt wording
\\[-2pt]
\noalign{\vskip 3pt}
\cdashline{1-3}
\noalign{\vskip 3pt}

\textbf{Relational transgression}
&
Illustrate behaviors or language that violate human relational norms \cite{yu-youth-ai-2026,yu-youthsafe-2025,chandra-2025,li-counselbench-2026,zhang-dark-side-ai-2025}
&
Emotional insensitivity and disregard; domination and manipulation; abrupt denial of service altogether; normalization of secrecy or coercion
\\[-2pt]
\noalign{\vskip 3pt}
\cdashline{1-3}
\noalign{\vskip 3pt}

\textbf{Limited clinical capabilities}
&
Have clinical limitations, yet may claim capabilities that an AI chatbot does not have \cite{cdt-dark-patterns,chandra-2025,jung-ai-mh-2025,zhang-dark-side-ai-2025,li-counselbench-2026,yu-youth-ai-2026}
&
Lacking clinical skills altogether (e.g., unable to respond appropriately to common mental health symptoms; things a ``good'' therapist would not do like stigmatize, collude with delusions, and enable suicidal ideation); marketing of unrealistic capabilities (e.g., stress management, medical advice)
\\

\midrule

\rowcolor{gray!20}
\multicolumn{3}{l}{
\rule[-6pt]{0pt}{18pt}
\textbf{AI chatbot's privacy promises don't reflect actual practices}
}\\
\addlinespace

\textbf{Privacy baiting}
&
Guide, ask, or coerce the user to share more information than needed \cite{cdt-dark-patterns,yu-youthsafe-2025}
&
Active prompting of more user info to ``complete'' an inquiry
\\[-2pt]
\noalign{\vskip 3pt}
\cdashline{1-3}
\noalign{\vskip 3pt}

\textbf{``Just between you and me''}
&
Give a pretense that information is private between the chatbot and the user, when in reality data can be used in various ways and is visible to companies \cite{cdt-dark-patterns}
&
``Whatever you share stays between you and me''
\\

\end{longtable}
\endgroup

\section{Positionality Statement}\label{app:positionality}
Given this study's focus on the mental health of young adults of color (YOC), we recognize that our identities, disciplinary training, and relationships to the topic shaped the research process. Our interdisciplinary team brought expertise in HCI, participatory design, behavioral social science, AI safety, clinical mental health, and public health, including prior experience working with YOC. One author is an undergraduate student.

Most authors identify as Asian or Asian American, with diverse national, ethnic, and immigration backgrounds. The first author is a first-generation Asian immigrant whose past experience with mental health and cultural background informed the study's conceptualization. At the same time, our experiences do not fully represent the situated knowledge of the Asian, Black, and Hispanic young adults who participated in this study, particularly across differences in age, race and ethnicity, culture, and life circumstances with rapidly emerging AI technologies. 

We acknowledge the power dynamics inherent in researcher-participant interactions during the data collection, especially among young adults. During data analysis, workshop facilitators iteratively discussed how our disciplinary backgrounds and prior experiences shaped theme development. We revisited workshop data many times in response to challenges to our interpretations from the wider team and retained participant quotations, where relevant, to foreground YOC perspectives and distinguish participants' accounts from our interpretations. 

\section{Pre- and Post-Participation Survey Data} \label{app:survey}

\begingroup
\setlength{\tabcolsep}{8pt}
\renewcommand{\arraystretch}{1.25}
\begin{center}
\begin{tabular}{@{}llcccc@{}}
\toprule
 & & \multicolumn{2}{c}{Pre-participation} & \multicolumn{2}{c}{Post-participation} \\
\cmidrule(lr){3-4}\cmidrule(lr){5-6}
Measure & Scale & Range & $M$ ($SD$) & Range & $M$ ($SD$) \\
\midrule
Trust in AI          & 1--5 & 2.00--5.00 & 3.33 (0.78) & 2.00--4.00 & 2.92 (0.51) \\
Perceived Quality    & 1--5 & 1.00--4.00 & 2.50 (1.38) & 1.00--4.00 & 2.08 (1.00) \\
General AI Attitude  & 1--5 & 1.90--4.10 & 3.23 (0.62) & 3.00--4.00 & 3.59 (0.40) \\
AI Self-Disclosure   & 1--7 & 1.50--6.25 & 3.79 (1.72) & 1.75--6.50 & 4.06 (1.71) \\
ChatGPT Literacy     & 1--5 & 3.40--4.60 & 4.03 (0.42) & 3.50--4.90 & 4.19 (0.43) \\
\bottomrule
\end{tabular}

\vspace{6pt}
\captionof{table}{Descriptive statistics for pre- and post-participation measures ($n = 12$)}
\label{tab:descriptives}
\end{center}
\endgroup

\vspace{4pt}
\noindent\small
\textit{Note.} $n = 12$. Sample includes 12 participants who completed both workshop days. Higher scores indicate more of the named construct. All values are rounded to two decimal places.

\smallskip
\noindent\textit{Trust in AI:} We asked one item (``How much do you trust the information or advice from AI chatbots?'') \cite{robb-common-sense-media-2025}.

\smallskip
\noindent\textit{Perceived Quality:} We asked one item (``How do conversations with AI chatbots compare to conversations with your real-life friends?'') \cite{robb-common-sense-media-2025}.

\smallskip
\noindent\textit{General Attitudes Toward AI:} We administered the Short General Attitudes Towards Artificial Intelligence Scale (Short GAAIS-10) \cite{schepman-gaais-2026}. The scale includes questions such as ``AI can have positive impacts on people's well-being'' and ``I think AI is dangerous'' (reverse-coded).

\smallskip
\noindent\textit{AI Self-Disclosure:} We included four self-disclosure questions about breadth (``I prefer to talk about myself to an AI chatbot'') and depth (``I disclose highly personal information to an AI chatbot'') \cite{cao-privacy-2026}. In the post-participation survey, we rephrased the question as their intended use going forward.

\smallskip
\noindent\textit{ChatGPT Literacy:} We included 10 questions from the ChatGPT literacy scale, with two questions per category---technical proficiency, critical evaluation, communication proficiency, creative application, and ethical competence \cite{lee-chatgpt-literacy-2024}. We selected an AI literacy scale about ChatGPT given the specific AI system used in the workshop.
\normalsize

\section{Co-design Outputs from Day 2}\label{app:design-outputs}
\begingroup
\small
\setlength{\tabcolsep}{6pt}
\renewcommand{\arraystretch}{1.1}

\newcommand{\DP}[1]{\text{#1.}}

\newcommand{\DesignPrincipleKey}{%
\begin{enumerate}[label=\arabic*., leftmargin=14pt]
    \setlength{\itemsep}{1pt}
    \setlength{\parskip}{0pt}
    \setlength{\parsep}{0pt}
    \item Adapt responses to users' identities and lived contexts
    \item Make self-disclosure feel structured and low-friction
    \item Uphold users' agency and autonomy in decision-making    
    \item Guide users' mindful, longitudinal engagement with AI
\end{enumerate}%
}

\begin{longtable}{
    >{\RaggedRight\arraybackslash}p{0.25\textwidth}
    >{\RaggedRight\arraybackslash}p{0.42\textwidth}
    >{\RaggedRight\arraybackslash}p{0.3\textwidth}
}
\caption{Co-design features from four teams and corresponding design principles}
\label{tab:co-design-output}\\

\toprule
\textbf{Features by participants} &
\textbf{Description} &
\textbf{Design principles by researchers} \\
\midrule
\endfirsthead

\multicolumn{3}{l}{
\textit{Table \thetable{} continued from previous page}
}\\
\toprule
\textbf{Features by participants} &
\textbf{Description} &
\textbf{Design principles by researchers} \\
\midrule
\endhead

\midrule
\multicolumn{3}{r}{\textit{Continued on next page}}\\
\endfoot

\bottomrule
\endlastfoot

\rowcolor{gray!20}
\multicolumn{3}{l}{
\rule[-6pt]{0pt}{18pt}
\textbf{Team 1: ``ChatGPT Wrapped'' style user report}
}\\
\addlinespace

\textbf{Usage summaries}
&
Daily/weekly recaps (e.g., time spent, topics covered) to support
pattern recognition in AI chatbot use
&
\multirow{3}{=}{%
    \DP{C} User agency and autonomy\par
    \DP{D} Mindful, longitudinal engagement with AI\par
}
\\[-2pt]
\noalign{\vskip 3pt}
\cdashline{1-2}
\noalign{\vskip 3pt}

\textbf{Youth-friendly \newline reflection tools}
&
Example features for self-reflection and peer discussions:
\begin{itemize}[leftmargin=12pt]
    \setlength{\itemsep}{1pt}
    \setlength{\parskip}{0pt}
    \setlength{\parsep}{0pt}
    \item ``Mood weather'' visualization of emotional patterns over time
    \item ``Archetypes'' capturing how their usage compares with other users
    \item ``ChatGPT wrapped'' style recap that can be shared with friends
\end{itemize}
\\[-2pt]
\noalign{\vskip 3pt}
\cdashline{1-2}
\noalign{\vskip 3pt}

\textbf{Embedded conversational nudges}
&
Gentle in-chat nudges that encourage breaks, self-monitoring, or social
reconnection, e.g.,
\begin{itemize}[leftmargin=*]
    \setlength{\itemsep}{1pt}
    \setlength{\parskip}{0pt}
    \setlength{\parsep}{0pt}
    \item ``Hey, we've been talking for 2 hours now. How about taking a break?''
    \item Small missions to reconnect with friends
\end{itemize}
&
\\

\midrule

\rowcolor{gray!20}
\multicolumn{3}{l}{
\rule[-6pt]{0pt}{18pt}
\textbf{Team 2: Intentional follow-up questions}
}\\
\addlinespace

\textbf{Identity-focused onboarding}
&
A pop-up questionnaire collects and saves background information on the
user, such as ethnicity, immigration history, geographic context, and
formative experiences, that inform future conversations and keep in mind the user's triggers and sensitive topics
&
\multirow{3}{=}{%
    \DP{A} Identity- and context-responsive interaction\par
    \DP{B} Structured, low-friction self-disclosure\par
    \DP{C} User agency and autonomy\par
    \DP{D} Mindful, longitudinal engagement with AI
}
\\[-2pt]
\noalign{\vskip 3pt}
\cdashline{1-2}
\noalign{\vskip 3pt}

\textbf{Adaptive re-prompting}
&
Similar questions are re-introduced later when the user discloses more
personal information in conversation
&
\\
\addlinespace
\noalign{\vskip 3pt}
\cdashline{1-2}
\noalign{\vskip 3pt}

\textbf{User-friendly response format}
&
Users answer intentionally curated questions in a survey-like format (e.g., buttons) that
avoids user fatigue and intrusiveness
&
\\[-2pt]
\noalign{\vskip 3pt}
\cdashline{1-2}
\noalign{\vskip 3pt}

\textbf{Clear opt-out option}
&
Users can opt out of answering to maintain privacy
&
\\

\midrule

\rowcolor{gray!20}
\multicolumn{3}{l}{
\rule[-6pt]{0pt}{18pt}
\textbf{Team 3: Conversational flow for harm reduction}
}\\
\addlinespace

\textbf{Mindful AI use reframing as ``harm reduction''}
&
Design for users who may face multi-level barriers (e.g., self-stigma, structural inaccess) to formal mental health support, with resources aimed at reducing further mental health harm
&
\multirow{3}{=}{%
    \DP{A} Identity- and context-responsive interaction\par
    \DP{C} User agency and autonomy\par
    \DP{D} Mindful, longitudinal engagement with AI
}
\\[-2pt]
\noalign{\vskip 3pt}
\cdashline{1-2}
\noalign{\vskip 3pt}

\textbf{Early risk identification}
&
Ability to detect and stratify mental health risk level early in the
conversation
&
\\[-2pt]
\noalign{\vskip 3pt}
\cdashline{1-2}
\noalign{\vskip 3pt}

\textbf{Non-intrusive follow-up}
&
Revisits of prior concerns later like a ``friend'' (e.g., ``did you go and
seek help by any chance?'')
&
\\[-2pt]
\noalign{\vskip 3pt}
\cdashline{1-2}
\noalign{\vskip 3pt}

\textbf{Role-based sub-chatbots}
&
Automatically activated sub-chatbots (within main chatbot) with different
roles based on topic, with confidentiality maintained across sub-chatbots
&
\\

\midrule

\rowcolor{gray!20}
\multicolumn{3}{l}{
\rule[-6pt]{0pt}{18pt}
\textbf{Team 4: Conversational flow for mutual accountability}
}\\
\addlinespace

\textbf{Overall design for accountability}
&
Chatbot that holds the user (more mindful about their AI usage) and itself
``accountable'' (acknowledging its capabilities and limitations)
&
\multirow{3}{=}{%
    \DP{C} User agency and autonomy\par
    \DP{D} Mindful, longitudinal engagement with AI
}
\\[-2pt]
\noalign{\vskip 3pt}
\cdashline{1-2}
\noalign{\vskip 3pt}

\textbf{User-controlled redirection}
&
Easily identifiable buttons for redirecting the conversation however the
user wants (e.g., ``do you want to connect with a mental health
professional? yes/no'')
&
\\[-2pt]
\noalign{\vskip 3pt}
\cdashline{1-2}
\noalign{\vskip 3pt}

\textbf{Features for usage awareness}
&
Built-in timer and periodic nudges to make chatbot use more visible and
encourage breaks
&
\\

\end{longtable}

\endgroup


\end{document}